\documentclass[twocolumn]{aastex631}
\usepackage{amsmath}

\newcommand{\fermi}{\textit{Fermi}}
\newcommand{\gr}{$\gamma$-ray}
\shorttitle{Finding Pulsar TeV Halos}
\shortauthors{Zheng \& Wang}

\begin{document}

\title{Finding Candidate TeV Halos among Very-High Energy Sources}

\author{Dong Zheng}
\affiliation{Department of Astronomy, School of Physics and Astronomy, Yunnan University, Kunming 650091, China; zhengdong@mail.ynu.edu.cn; wangzx20@ynu.edu.cn}

\author{Zhongxiang Wang}
\affiliation{Department of Astronomy, School of Physics and Astronomy, Yunnan University, Kunming 650091, China; zhengdong@mail.ynu.edu.cn; wangzx20@ynu.edu.cn}
\affiliation{Key Laboratory for Research in Galaxies and Cosmology, Shanghai Astronomical Observatory, Chinese Academy of Sciences, 80 Nandan Road, Shanghai 200030, China}



\begin{abstract}
	We search for possible pulsar TeV halos among the very-high-energy 
	(VHE) sources reported in different VHE surveys, among which 
	in particular we use the results from the first Large High Altitude 
	Air Shower Observatory 
	(LHAASO) catalog of $\gamma$-ray sources. Six candidates are found.
	They share similar properties of containing a middle-aged, 
	\gr--bright 
	pulsar in their positional error circles (the respective pulsars are 
	J0248+6021, J0359+5414, J0622+3749, J0633+0632, J2006+3102, and 
	J2238+5903), being in a rather clean 
	field without any common Galactic VHE-emitting supernova remnants 
	or (bright) pulsar wind nebulae (PWNe), and showing an
	absence of any 
	\gr\ emissions in 0.1--500\,GeV after removing the pulsars' emissions.
	Combining these candidates with several reported
	(candidate) TeV halos,
	we obtain the relationships between their luminosity
	at 50\,TeV ($L_{\rm 50TeV}$) and the corresponding
	pulsars' spin-down energy ($\dot{E}$), which are 
	$L_{\rm 50TeV}\sim \dot{E}^{0.9}$ and 
	$L_{\rm 50TeV}/\dot{E}\sim 6.4\times 10^{-4}$. The relationships
	are nearly identical to previously reported ones. We probe possible
	connections between the extension sizes of the VHE sources and
	the pulsars' ages, and find a weak older-and-smaller trend.
	By comparing to the VHE detection results for PWNe, it is clear 
	that the (candidate) TeV halos have hard emissions by either
	having their power-law indices be smaller than 2 in 1--25\,TeV 
	or by only being detected in 25--100\,TeV. In addition, we 
	also consider seven other VHE sources as possible TeV halos 
	based on the results from different studies of them,
	but they do not fit cleanly with the properties listed above, 
	indicating their potentially complex nature.
\end{abstract}



\section{Introduction}
\label{sec:intro}

TeV halos are extended Very-High-Energy (VHE; $\geq$100\,GeV) $\gamma$-ray 
emissions around middle-aged pulsars ($\sim$100\,kyr). Their existence
has been firmly established due to the detection of extended emissions
around the nearby 
pulsars Geminga and Monogem, as observed by the High-Altitude Water Cherenkov 
(HAWC) 
Observatory \citep{abe+17}. Since the detection revelation, various studies 
focusing
on their general existence and possible properties have been carried out
(see, e.g., \citealt{lab+17}, \citealt{slb19}, \citealt{fan22}, 
and \citealt{ml22} and references therein).
Importantly, TeV halos can be a significant contributor of cosmic electrons and 
positrons in our Galaxy \citep{gml+20, lwa+22, ylz+24}. 

\begin{deluxetable*}{ccccccccccc}
\tablecaption{Properties of pulsars and likely associated VHE sources.
\label{tab:psr}}
\tablewidth{0pt}
\tablehead{
\colhead{1LHAASO} & \colhead{$P_{0}$} & \colhead{$\dot{P}$}  & \colhead{$\dot{E}/10^{35}$} & \colhead{Distance} & \colhead{Age} & \colhead{$F^{\rm PSR}_{\rm X-ray}/10^{-13}$}& \colhead{$F^{\rm PWN}_{\rm X-ray}/10^{-13}$}& \colhead{$F_{\rm 50TeV}/10^{-13}$}& \colhead{Extension} & \colhead{References} \\
\colhead{PSR}            & \colhead{(s)}         & \colhead{($10^{-14}$)} & \colhead{(erg s$^{-1}$)}     & \colhead{(kpc)}       & \colhead{(kyr)} & \colhead{(erg cm$^{-2}$ s$^{-1}$)}& \colhead{(erg cm$^{-2}$ s$^{-1}$)}& \colhead{(erg cm$^{-2}$ s$^{-1}$)}& \colhead{(deg)}& \colhead{} }
\startdata
        J0249$+$6022  &   &  &   &   &  &  &   &3.72  $\pm$ 0.36    & 0.38 $\pm$ 0.08&  \\
        J0248$+$6021  & 0.22 & 5.51 & 2.13 & $2.0 \pm 0.2$         & 62.4  & $<$ 9.0             & --                                   & &  & 1, 2 \\
        \hline
	J0359$+$5406  &   &  &   &   &  &  &   &3.40  $\pm$ 0.24    &0.30 $\pm$ 0.04&       \\
	J0359$+$5414  & 0.08 & 1.67 & 13.0 & 3.45                     & 75.2  & 0.09 $\pm$ 0.03 & 0.20 $\pm$ 0.03         & &  &  3    \\
	\hline
	J0622$+$3754  &   &  &   &   &  &  &   &5.68  $\pm$ 0.28    &0.46 $\pm$ 0.03 &  \\
	J0622$+$3749  & 0.33 & 2.54 & 0.27 & $<$3.47               & 208   & $<$0.14           & --                                    & &  & 4  \\
	\hline
	J0635$+$0619  &   &  &   &   &  &  &   &3.76  $\pm$ 0.40    &0.60 $\pm$ 0.07&     \\
	J0633$+$0632  & 0.30 & 7.96 & 1.20 & 1.35$^{+0.65}_{-0.65}$  & 59.2  & $0.33 \pm 0.06$   & $1.17^{+0.11}_{-0.13}$  & &  &  5    \\
	\hline
	J2005$+$3050  &   &  &   &   &  &  &   &1.84  $\pm$ 0.20    &0.27 $\pm$ 0.05&    \\
	J2006$+$3102  & 0.16 & 2.49 & 2.24 & 4.7                       & 104   & $<$ 9.0             & --                                   & &  &  6   \\
	\hline
	J2238$+$5900  &   &  &   &   &  &  &   &8.12  $\pm$ 0.48    &0.43 $\pm$ 0.03 &    \\
	J2238$+$5903  & 0.16 & 9.70 & 8.89 & 2.83                     & 26.6  & $<$ 0.44           & --                                   & &  &  4  \\
	\hline
	\hline
	J0542$+$2311u  &   &  &   &   &  &  &   &11.72 $\pm$ 0.48    &0.98 $\pm$ 0.05     &             \\
	B0540$+$23  & 0.25  & 1.54 & 0.41 & 1.57    & 253   & 0.08 $\pm$ 0.04
& --  & &  &  4            \\ \hline
	J1740$+$0948u   &   &  &   &   &  &  &   &1.64 $\pm$ 0.16    & $<$ 0.11 &                      \\
	J1740$+$1000   & 0.15  & 2.13 & 2.32 & 1.23    & 114   & 0.24 $\pm$ 0.02 & 0.60 $\pm$ 0.06  & &  &  7, 8                     \\
	\hline
	J1809-1918u      &   &  &   &   &  &  &   &37.84 $\pm$ 5.08    & $<$ 0.22 &      \\
	J1809-1917      & 0.08  & 2.55 & 17.8 & 3.27    & 51.4  & 0.47$^{+0.01}_{-0.04}$ & 2.6--4.9  & &  &  9     \\
	\hline
	J1813-1245     &   &  &   &   &  &  &   &5.68 $\pm$ 1.08    & $<$ 0.31 &                     \\
	J1813-1246     & 0.05  & 1.76 & 62.4 & 2.64    & 43.4  & 10.80 $\pm$ 0.10 & $<$ 1.5 & &  &   10                    \\
	\hline
	J1825-1256u   &   &  &   &   &  &  &   &20.32 $\pm$ 1.68   & $<$ 0.2 & \\
	J1826-1256   & 0.11 & 12.1 & 36.0 & 1.55         & 14.4  & $1.04^{+0.14}_{-0.13}$ & $0.85^{+0.10}_{-0.09}$ & &  &     11                  \\
	\hline
	J1825-1337u      &   &  &   &   &  &  &   &40.40$\pm$ 2.44   & $<$ 0.18 &                     \\
	J1826-1334      & 0.10  & 7.53 & 28.4 & 3.61    & 21.4  & 0.16 $\pm$ 0.04 & 4.5$^{+0.3}_{-0.2}$ & &  &  12                     \\
	\hline
	\hline
	J1928$+$1746u &   &  &   &   &  &  &   &2.88 $\pm$ 0.28   & $<$ 0.16 & \\
	J1928$+$1813u &   &  &   &   &  &  &   &9.92 $\pm$ 0.64   & 0.63 $\pm$ 0.03 &                     \\
	J1928$+$1746 & 0.07  & 1.32 & 16.0 & 4.34    & 82.6  & $<$ 0.08 & --  & &  &          13             \\
\enddata
\tablecomments{References for X-ray fluxes and distances: (1) \citet{mdc11}, (2) \citet{tpc+11}, (3) \citet{zks18}, (4) \citet{pb15}, (5) \citet{dko+20}, (7) \citet{nab+13}, (7) \citet{rma+22}, (8) \citet{kmp+08}, (9) \citet{kyh+20}, (10) \citet{mhp+14}, (11) \citet{kzs19}, (12) \citet{pkb08}, (13) \citet{kdp+12}}
\end{deluxetable*}

In our recent studies of Galactic VHE sources for those whose nature is not 
clear (i.e., unidentified), we have focused on finding and studying 
their possible lower energy counterparts by analyzing available
multi-energy data (e.g., \citealt{xin+22,zhe+23}).
The primary ones used
are the GeV \gr\ data (in energy range of 0.1--500\,GeV) obtained with 
the Large Area Telescope (LAT) onboard
the {\it Fermi Gamma-ray Space Telescope (Fermi)}. It has been ascertained that
for a significant fraction of VHE sources, a known pulsar is often found 
located in the field, within the error circle of such a VHE source 
(e.g., \citealt{3hwc}). Some of
these pulsars with the positional coincidence are \gr-bright, which can cause 
difficulties in analyses because of the low-spatial resolution of the LAT
data (e.g., $\sim$1\,deg at 1\,GeV). The strategy we have applied to overcome 
such difficulties is to remove
the `contamination' of a pulsar, the pulsed emission, by timing this pulsar 
at $\gamma$-rays. This helps reveal the residual emission in a source 
field, which allows us to conduct clean studies of it as a candidate 
counterpart (e.g., \citealt{xin+22}).

However in the studies of the VHE sources 3HWC~J0631+107
(\citealt{3hwc}; or 1LHAASO J0631+1040), 1LHAASO J1959+2846u, and
1LHAASO J2028+3352, where 1LHAASO stands for the first Large High Altitude Air
Shower Observatory (LHAASO; \citealt{cdl+19}) catalog of Gamma-ray sources
\citep{1lhaaso}, no significant residual emissions were found
after we removed the pulsed emissions of PSRs~J0631+1036, J1958+2846, 
and J2028+3332, respectively. The non-detections, combined with 
the pulsars' many similarities to Geminga and the fact
that no primary Galactic VHE-emitting sources (e.g., \citealt{hgps}),
such as supernova remnants (SNRs) or pulsar wind nebulae (PWNe), 
are known in the fields, led to our identification
of the three VHE sources as being TeV halos powered by their respective pulsars
\citep{zwx23,zw23}.

\begin{deluxetable*}{cccccc}[htbp]
\tablecaption{Timing solutions and phase ranges for six pulsar targets.
\label{tab:Timing}}
\tablewidth{0pt}
\tablehead{
\colhead{Source} & \colhead{End time} & \colhead{$f$} & \colhead{$f_1$/$10^{-12}$} & \colhead{On-pulse} & \colhead{Off-pulse}  \\
\colhead{}            & \colhead{(MJD)} & \colhead{(Hz)} & \colhead{(Hz s$^{-1}$)} & \colhead{} & \colhead{} 
}
\startdata
                     J0248$+$6021 & 58839&4.605826479 & $-$1.17136   & 0.125--0.5625            &0--0.125, 0.5625--1   \\
                     J0359$+$5414 & 58044&12.58999189 & $-$2.64978   & 0.125--0.5625            & 0--0.125, 0.5625--1  \\
                     J0622$+$3749 & 58835&3.001119808  & $-$0.228945  & 0--0.625, 0.9375--1     & 0.625--0.9375        \\
                     J0633$+$0632 & 58738&3.362415888 & $-$0.899654 & 0--0.3125, 0.5--0.625   & 0.3125--0.5, 0.625--1   \\
                     J2006$+$3102 & 57697&6.108786932 &$-$0.928096 & 0.375--0.625              & 0--0.375, 0.625--1   \\
                     J2238$+$5903 & 58680&6.144764381 & $-$3.65692   & 0--0.375, 0.5--0.625     &0.375--0.5, 0.625--1   \\
\enddata
\tablecomments{Frequencies were from 3PC.}
\end{deluxetable*}

Our series of work and obtained results suggest that there could be more TeV 
halos
among the unidentified VHE sources. Particularly for those reported in 1LHAASO,
the energy range spreads from 1\,TeV to approximately 100\,TeV, which is
covered by
the LHAASO Water Cherenkov Detector Array (WCDA) in 1--25\,TeV and Kilometer 
Square Array (KM2A) in 25--100\,TeV \citep{cdl+19}. 
The wide energy-range coverage allows us to 
select sources with spectra harder than those of the PWNe \citep{haa+18}, which
is a possible feature that may be used to differentiate the TeV halos from 
the PWNe
(see \citealt{zwx23} and \citealt{zw23}). In addition, the Third \fermi\ Large 
Area Telescope Catalog of Gamma-ray Pulsars (3PC) has very recently been 
released \citep{3pc}. It provides the timing solutions for 
\fermi-LAT--detected \gr\ pulsars. These solutions allow us to easily carry out
our studies of VHE sources when pulsed \gr\ emissions need to be removed. 
We have thus conducted
a further search for candidate TeV halos. We have found six good candidates
(first six in Table~\ref{tab:psr}) and report the results in this paper.

In this work, we mainly used the detection results in 1LHAASO, but 
also included results reported in the High Energy 
Spectroscopy System (HESS) Galactic plane survey (HGPS; \citealt{hgps}) and
in the third HAWC catalog (3HWC; \citealt{3hwc}). In some cases, reported
results from the observations conducted with
the Very Energetic Radiation Imaging Telescope Array System (VERITAS) 
and the Milagro Gamma-ray Observatory (MGRO; \citealt{aaab+09})
were also used. To be as complete as possible, we essentially went through 
all the Galactic VHE sources that are likely associated with a pulsar
and show some aspects of a TeV halo. As a result, we found another seven 
sources and list
them in the lower part of Table~\ref{tab:psr}. Some of these sources are 
in a complex region, such as being potentially associated with an SNR/PWN 
in the field, and some contain a pulsar that does not have \gr\ emission or 
clear off-pulse phases in the case of being \gr\ bright. We included
these sources in our discussion (Section~\ref{sec:dis}), and a
brief introduction for each of them
is provided in Appendix~\ref{sec:src}.

In the following Section~\ref{sec:da}, we describe the \fermi-LAT data 
we used and our data analyses, which include how we obtained
the off-pulse data of the six pulsar targets through pulsar timing. 
In Section~\ref{sec:res}, in conjunction with our analysis
results, we provide the properties of each pulsar target and its associated
VHE source, which helps identify the latter as a candidate
TeV halo. In Section~\ref{sec:dis}, we discuss these 
sources' general properties by considering the VHE sources as being TeV halos
powered by the corresponding pulsars.

\section{Data Analysis}
\label{sec:da}

\subsection{LAT Data and Source Model}
\label{sec:data}

Photon data files with timing analysis results for each of
the \fermi-LAT--detected pulsars are provided in 3PC. There are two types 
of data with different sizes, one containing photons within $3^{\circ}$ 
with an energy band ranging from 50\,MeV to 300\,GeV\footnote{https://heasarc.gsfc.nasa.gov/FTP/fermi/data/lat/catalogs\\/3PC/photon/3deg\_50MeV/}, 
and the other containing photons of energies from
20\,MeV to 1\,TeV within $15^{\circ}$\footnote{https://heasarc.gsfc.nasa.gov/FTP/fermi/data/lat/catalogs\\/3PC/photon/15deg\_20MeV/}. Both types 
of data files 
are centered at the position of each pulsar. The data were
selected from the latest \fermi\ Pass 8 database with Event Class of 128. 
Events with a zenith angle larger than $105^{\circ}$ and bad quality flags
were excluded. We used both of the data files in the following analyses.

\begin{figure*}
\gridline{\fig{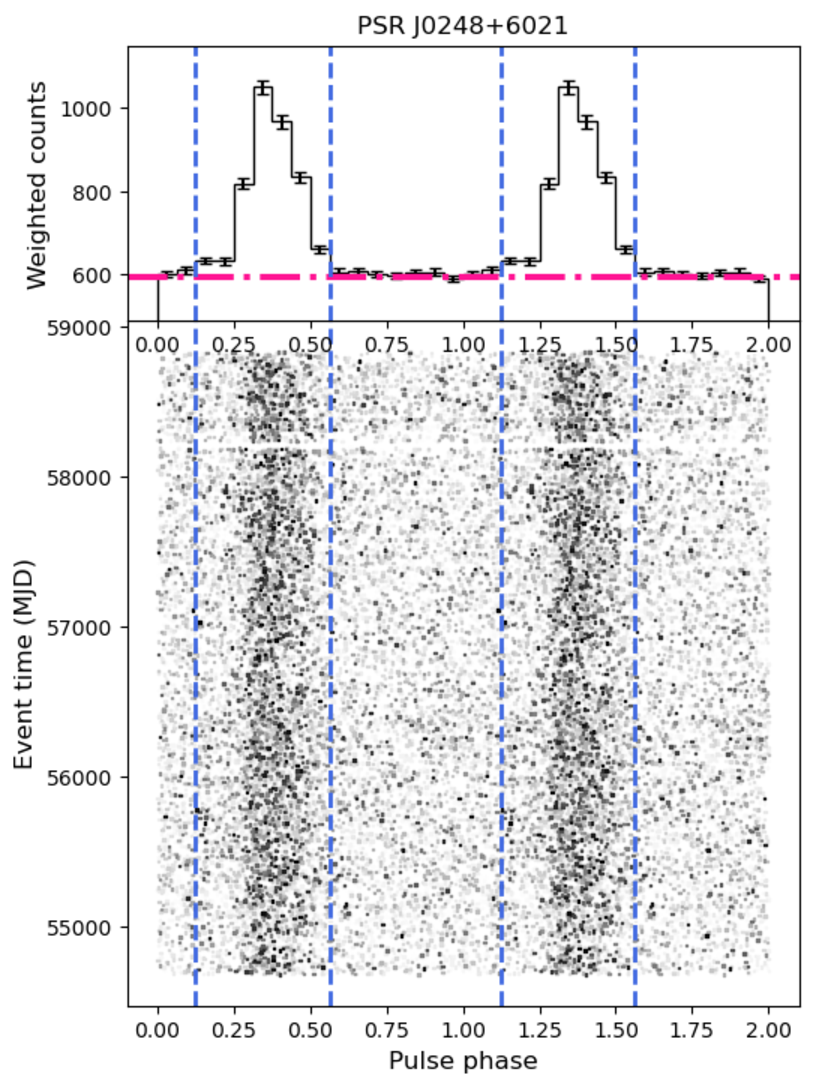}{0.3\textwidth}{}\hspace{-15mm}
          \fig{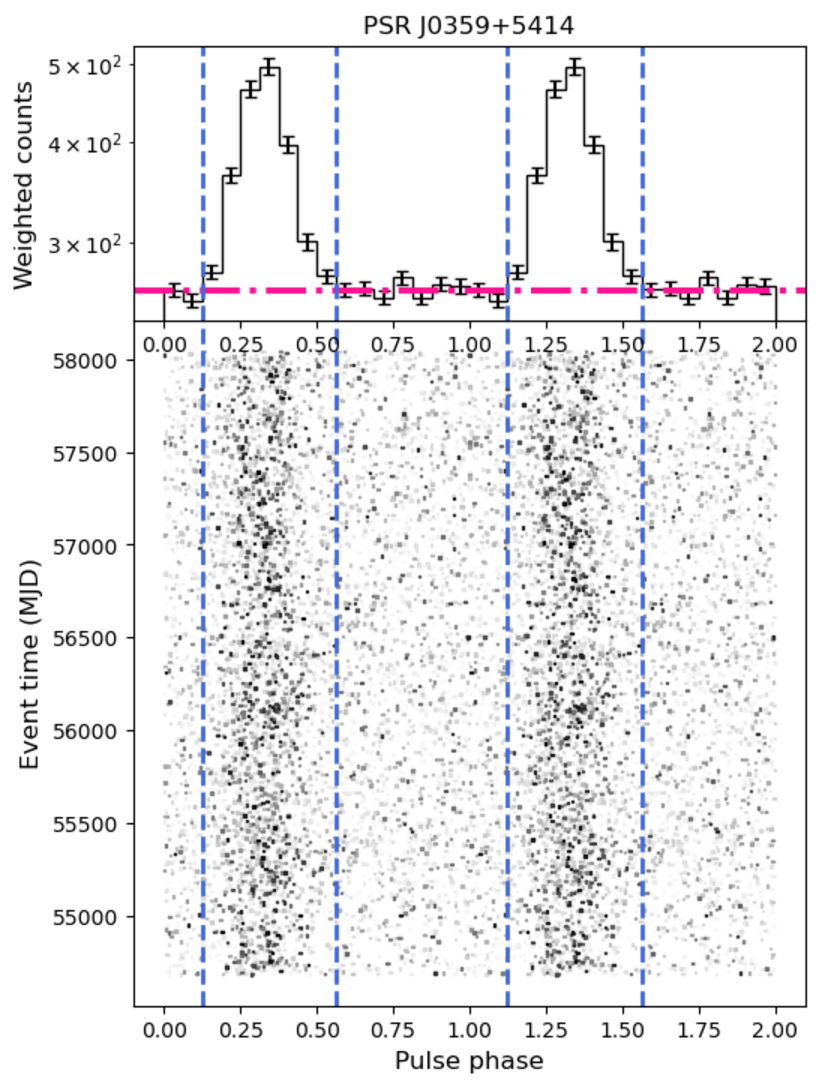}{0.3\textwidth}{}\hspace{-15mm}
          \fig{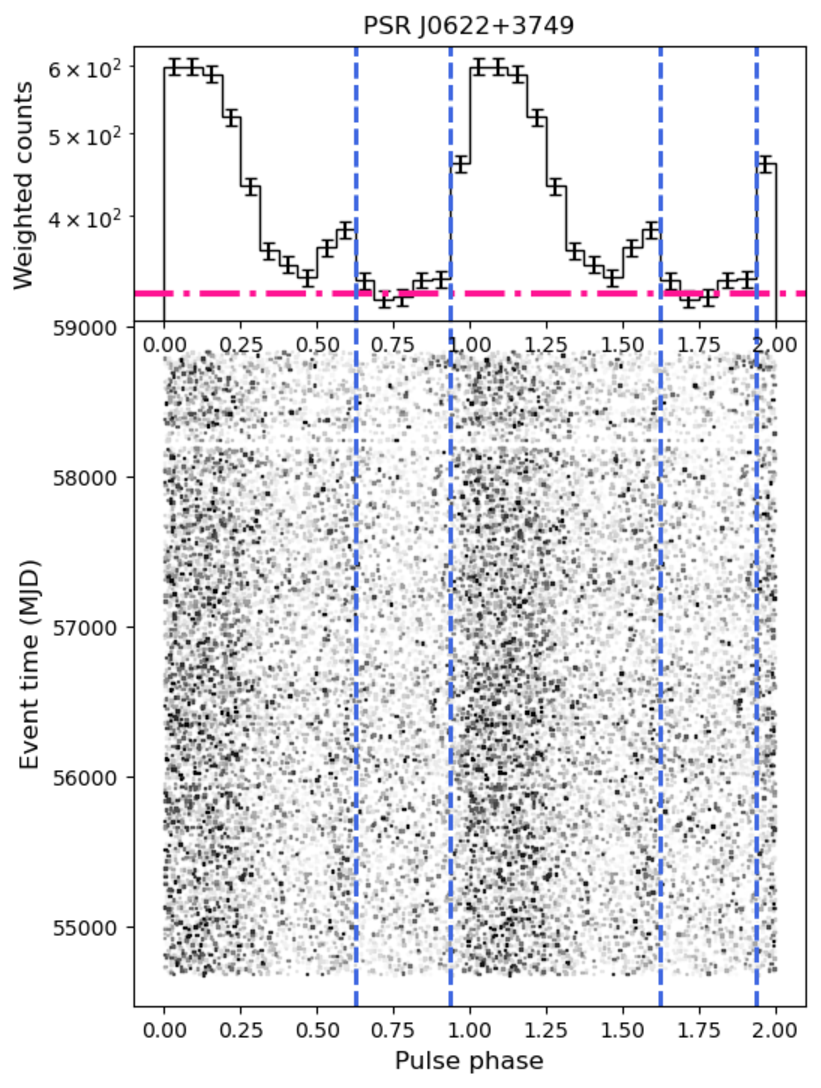}{0.3\textwidth}{}
          }\vspace{-7mm} 
          
\gridline{\fig{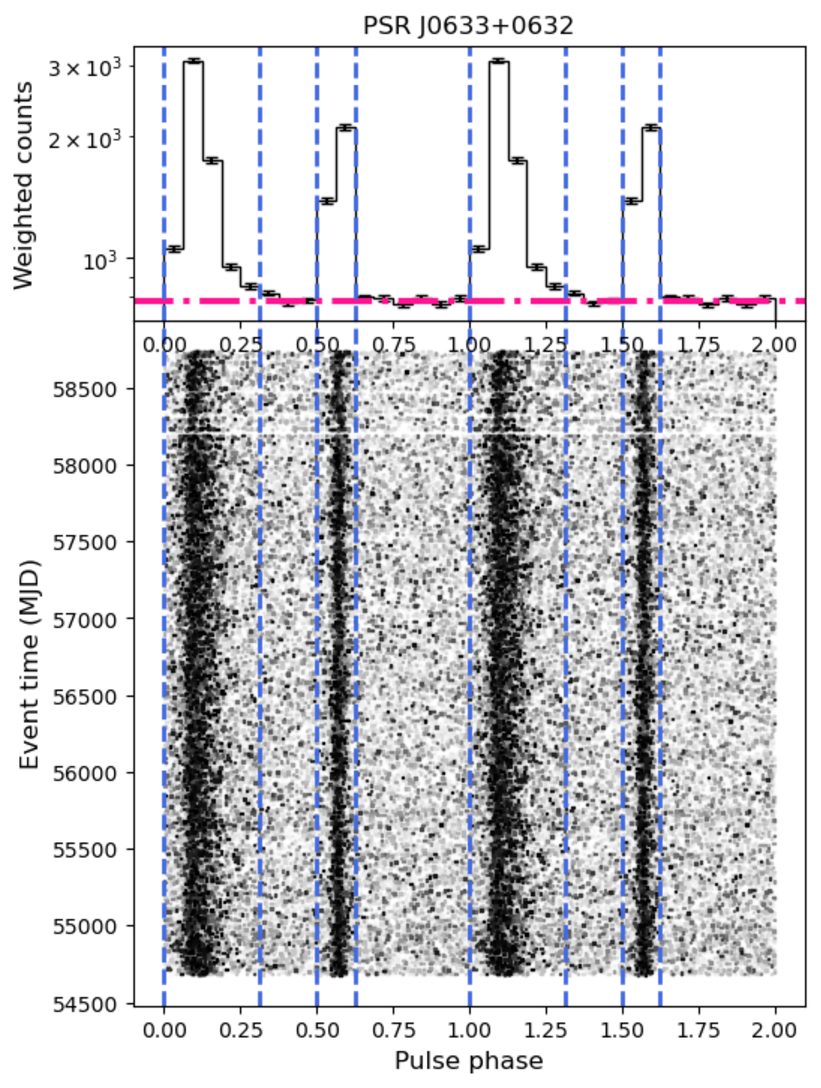}{0.3\textwidth}{}\hspace{-15mm}
          \fig{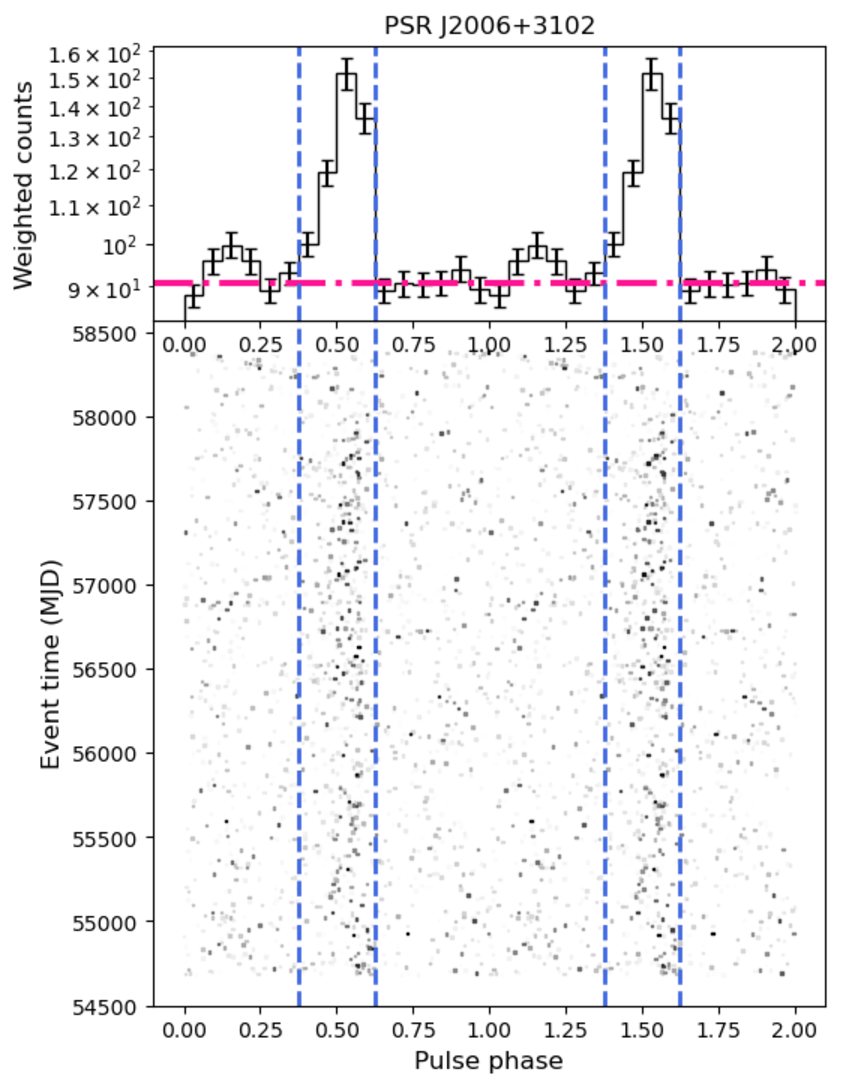}{0.3\textwidth}{}\hspace{-15mm}
          \fig{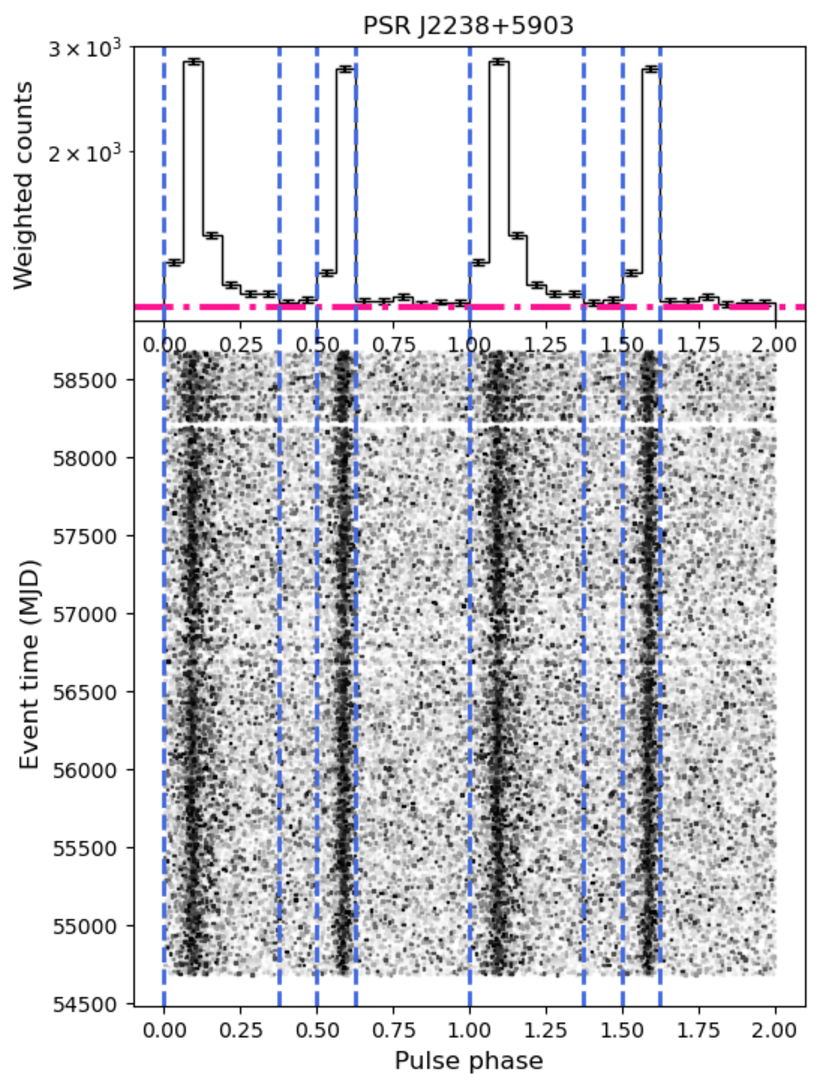}{0.3\textwidth}{}
          }\vspace{-7mm} 
\caption{Pulse profiles (top) and two-dimensional phaseograms (bottom) of
six pulsar targets. The on- and off-pulse phase ranges we defined based
	on the pulse profiles are marked by the dashed lines. 
\label{fig:phase}}
\end{figure*}

In our analyses, we chose photons in the energy band of 0.1--500\,GeV 
with a zenith angle of $<90^{\circ}$. 
The start times of the data are 
2008 August 4 15:43:36 (UTC), but because the timing solutions did not cover 
the whole observation time period of \fermi-LAT, the end times are different
and are given in Table~\ref{tab:Timing} for each of the pulsar targets. 
We set the regions of interests (RoIs) with a size of 
$15^{\circ} \times 15^{\circ}$, centered at each of the pulsar 
targets. The latest \fermi-LAT Fourth Source Catalog  (4FGL-DR4, \citealt{dr4})
was used to construct source models. For each target,
the sources within a range of 15$^{\circ}$ radius were included,
and their 4FGL-DR4 spectral forms were used. In addition,
two background models, the Galactic and extragalactic diffuse emission, were 
included in the source models, which were the files gll\_iem\_v07.fits and 
iso\_P8R3\_SOURCE\_V3\_v1.txt, respectively.

\subsection{Timing Analysis}
\label{sec:timing}

In both the $3^{\circ}$ and $15^{\circ}$ photon data files, 
a photon's 
probability (to be from a pulsar) and spin phase (of the pulsar) were 
included. We used the $3^{\circ}$ photon files to construct the pulse profiles
and define the on- and 
off-pulse phase ranges. The pulse profiles, with the on- and off-pulse
phase ranges marked, are shown in Figure \ref{fig:phase}. The values
of the phase ranges, as well as the timing solutions, are given
in Table~\ref{tab:Timing}. 

\begin{deluxetable*}{ccccccc}
\tablecaption{Binned likelihood analysis results from the on- and off-pulse 
	phase data.
\label{tab:likelihood}}
\tablewidth{0pt}
\tablehead{
\colhead{Pulsar} & \colhead{Phase Range} & \colhead{$F_{0.1-500}/10^{-8} $}           & \colhead{$\Gamma$} & \colhead{ ExpfactorS} & \colhead{TS}  \\
\colhead{}            & \colhead{}                       & \colhead{(photons s$^{-1}$ cm$^{-2}$)} & \colhead{}                   & \colhead{} & \colhead{} 
}
\startdata
         J0248$+$6021     & On-pulse &$3.91 \pm 0.27$     & $2.32 \pm 0.04$  & $ 0.89 \pm 0.08$ & 2828.9   \\
	                              & Off-pulse & $\leq$0.11                        & 2                          & -                           &  0.2 \\                        
	\hline
	 J0359$+$5414     & On-pulse &$2.77 \pm 0.20$     & $2.21 \pm 0.04$ & $ 0.52 \pm 0.07$ & 1658.3   \\
	                              & Off-pulse & $\leq$0.03                       & 2                              & -                           &  0.0 \\      
	\hline
	 J0622$+$3749     & On-pulse &$2.56 \pm 0.14$     & $2.36 \pm 0.04$  & $ 1.19 \pm 0.10$ & 2686.7   \\
	                              & Off-pulse & $\leq$0.08                        & 2                          & -                           &  1.5 \\                        
	\hline
	J0633$+$0632      & On-pulse &$8.49 \pm 0.32$     & $1.97 \pm 0.02$ & $ 0.63 \pm 0.03$ & 18803.2   \\
	                              & Off-pulse & $\leq$0.06                       & 2                              & -                           &  0.0 \\   
	\hline
	 J2006$+$3102     & On-pulse &$1.06 \pm 0.18$     & $2.15 \pm 0.08$  & $ 0.65 \pm 0.14$ & 424.0   \\
	                              & Off-pulse & $\leq$0.19                       & 2                              & -                           &  1.2 \\   
	\hline
	 J2238$+$5903      & On-pulse &$8.24 \pm 0.30$     & $2.25 \pm 0.02$  & $ 0.51 \pm 0.03$ & 8913.0   \\
	                              & Off-pulse & $\leq$0.17                        & 2                              & -                           &  4.3 \\   
	\hline
\enddata
\end{deluxetable*}

\subsection{Likelihood Analysis of the on- and off-pulse Data}
\label{sec:la}

\subsubsection{On-pulse Data}
\label{sec:Onpulse}

The standard binned likelihood analysis was performed to the on-pulse data
of each pulsar in 0.1--500\,GeV. The spectral parameters of sources 
in a source model within $5^{\circ}$ from a pulsar target were set as free, 
while 
those of the other sources were fixed at the values given in 4FGL-DR4. 
In addition, the normalizations of the two background components were set 
as free parameters. 
We used a PLSuperExpCutoff4 (PLSEC; \citealt{4fgl-dr3}) model shape to fit 
the on-pulse data
of the pulsars. There are two forms of PLSEC based on the conditions set for
the forms (see \citealt{4fgl-dr3} for details). 
One, $\frac{dN}{dE} = N_{0} (\frac{E}{E_{0}})^{- \Gamma - \frac{d}{2} ln(\frac{E}{E_{0}}) - \frac{db}{6} ln^{2}(\frac{E}{E_{0}}) - \frac{db^{2}}{24} ln^{3} (\frac{E}{E_{0}})}$,
was used for PSRs J0359+5414, J0633+0632, J2006+3102 and J2238+5903
(hereafter J0359, J0633, J2006, and J2238, respectively), and the other,
$\frac{dN}{dE} = N_{0} (\frac{E}{E_{0}})^{- \Gamma + \frac{d}{b} } \exp \{\frac{d}{b^{2}} [1 - (\frac{E}{E_{0}})^{b}]\}$, was used for PSRs~J0248+6021 and 
J0622+3749 (hereafter J0248 and J0622, respectively).
In the two forms, $\Gamma$ and $d$ (or {\tt ExpfactorS}) are the photon index 
and the local curvature at energy $E_{0}$, respectively, and $b$ is a measure 
of the shape of the exponential 
cutoff. Following 4FGL-DR4, we fixed the value of $b$ at $2/3$ for
our analysis. 
The likelihood analysis results for each pulsar are given in 
Table \ref{tab:likelihood}.

\begin{figure*}
\gridline{\fig{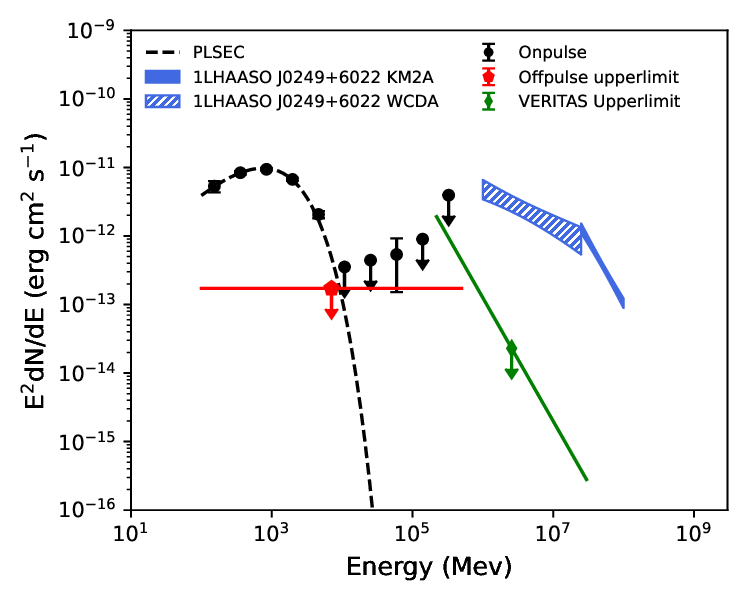}{0.33\textwidth}{(a) PSR J0248$+$6021}\hspace{-10mm}
          \fig{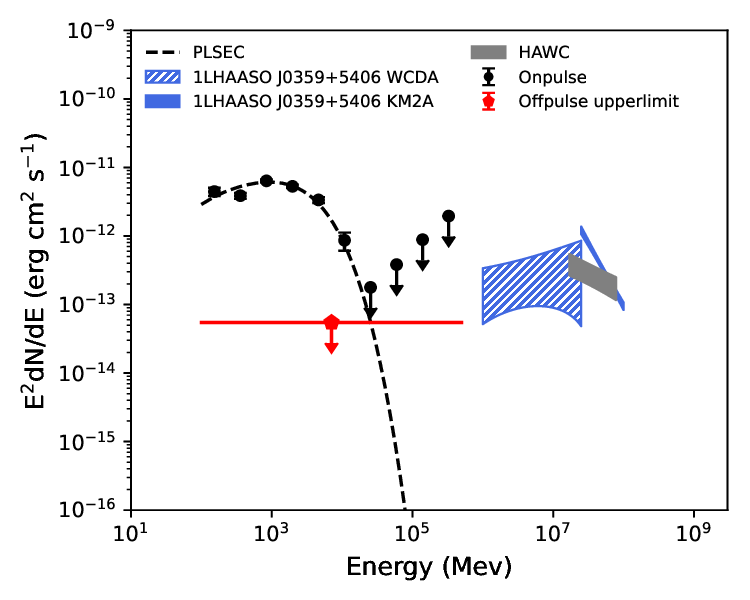}{0.33\textwidth}{(b) PSR J0359$+$5414}\hspace{-10mm}
          \fig{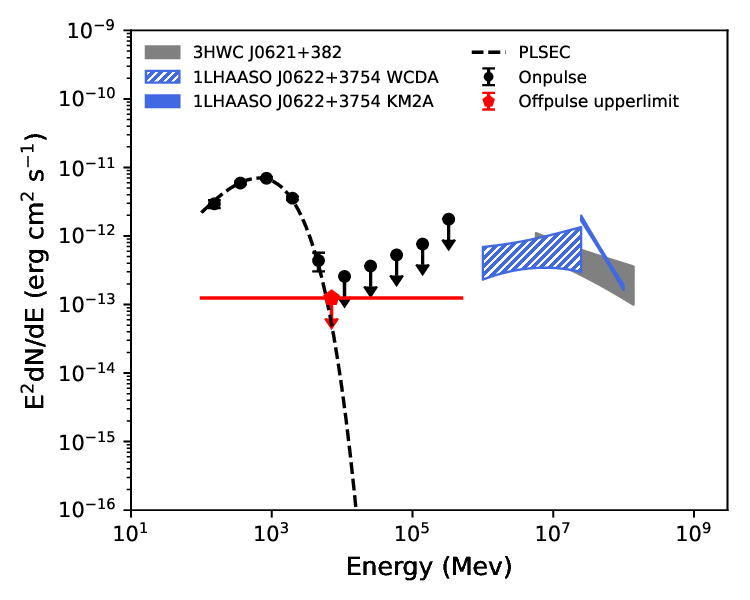}{0.33\textwidth}{(c) PSR J0622$+$3749}
          }
\gridline{\fig{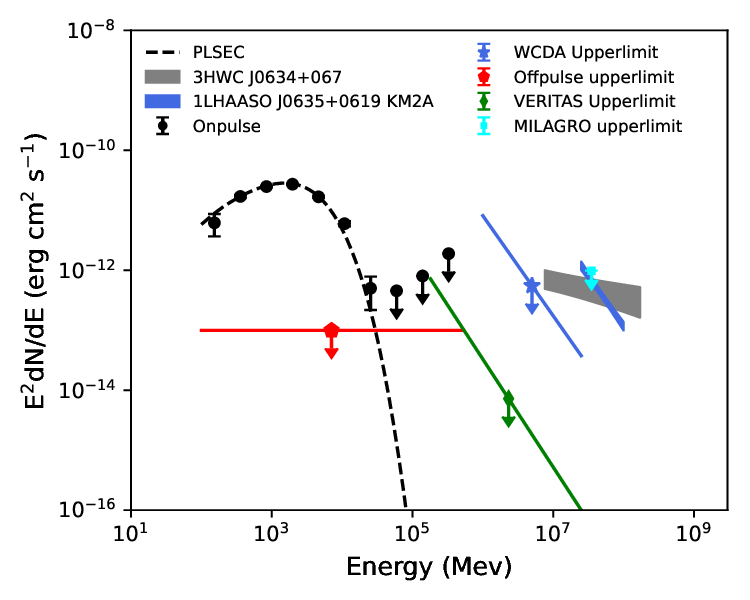}{0.33\textwidth}{(d) PSR J0633$+$0632 }\hspace{-10mm}
              \fig{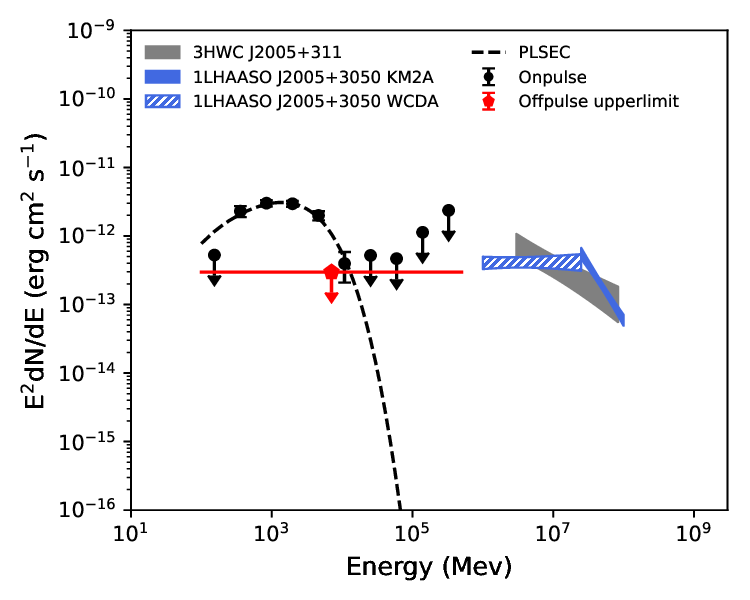}{0.33\textwidth}{(e) PSR J2006$+$3102}\hspace{-10mm}
             \fig{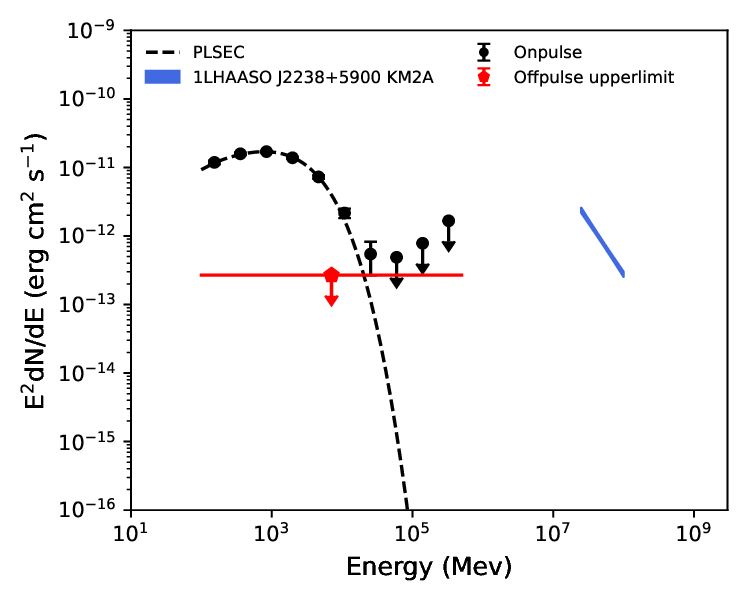}{0.33\textwidth}{(f) PSR J2238$+$5903}
             }
 \caption{Spectra and spectral upper limits of the six pulsars
	during their on- and off-pulse phase ranges, which are shown as 
	black data points 
	(and black dashed curves, the best-fit PLSEC models) 
	and red lines (assuming $\Gamma = 2$),
	respectively. In addition, we overplot the spectra of the LHAASO 
	sources, and available HAWC flux measurements and/or 
	VERITAS \citep{abb+19} and MGRO \citep{aaab+09} flux upper limits of
	the VHE sources. For details, see Section~\ref{sec:res} and 
	Figure~\ref{fig:tsmap}.
\label{fig:spec}}
\end{figure*}

We also obtained the spectral data points of the pulsars from their on-pulse 
data.
The energy range from 0.1 to 500\,GeV was evenly divided logarithmically 
into 10 bins. Binned likelihood analysis was performed to each bin's data 
to obtain the fluxes.
In this analysis, the normalizations of the sources within $5^{\circ}$ 
of a pulsar and the two background components were set as free parameters, 
while the other parameters were fixed at the values obtained above from
the binned likelihood analysis of the data in the whole energy range. 
When the test statistic (TS) value of a bin was $<$4, we replaced the flux with
the 95\% upper limit as derived from the data of the bin. The obtained spectra
are shown in Figure~\ref{fig:spec}.

\subsubsection{Off-pulse Data}
\label{sec:Offpulse}

We also performed standard binned likelihood analysis to the off-pulse data 
of the pulsars. The parameter setup was the same as that in the above analysis 
of the on-pulse data (Section~\ref{sec:Onpulse}). We assumed a power law 
(PL) for any emission at the position of each pulsar, 
$\frac{dN}{dE} = N_{0} (\frac{E}{E_{0}})^{-\Gamma}$. From the analysis,
no significant emissions were detected during the off-pulse phase ranges of
the pulsars.  In Table~\ref{tab:likelihood},
we provided the TS values when we assumed $\Gamma = 2$ as the exemplary result.

\begin{figure*}
\gridline{\fig{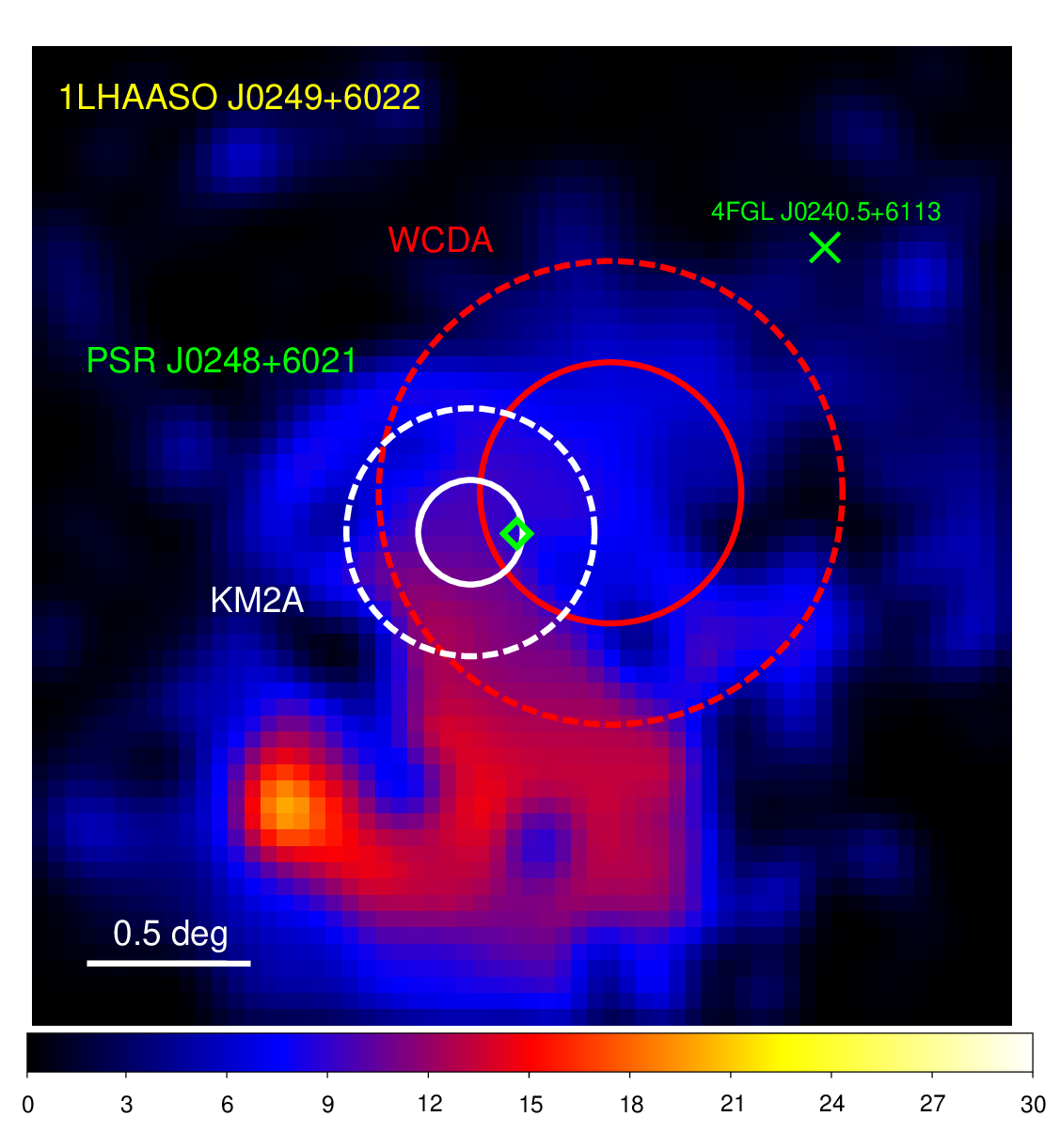}{0.33\textwidth}{(a)}
          \fig{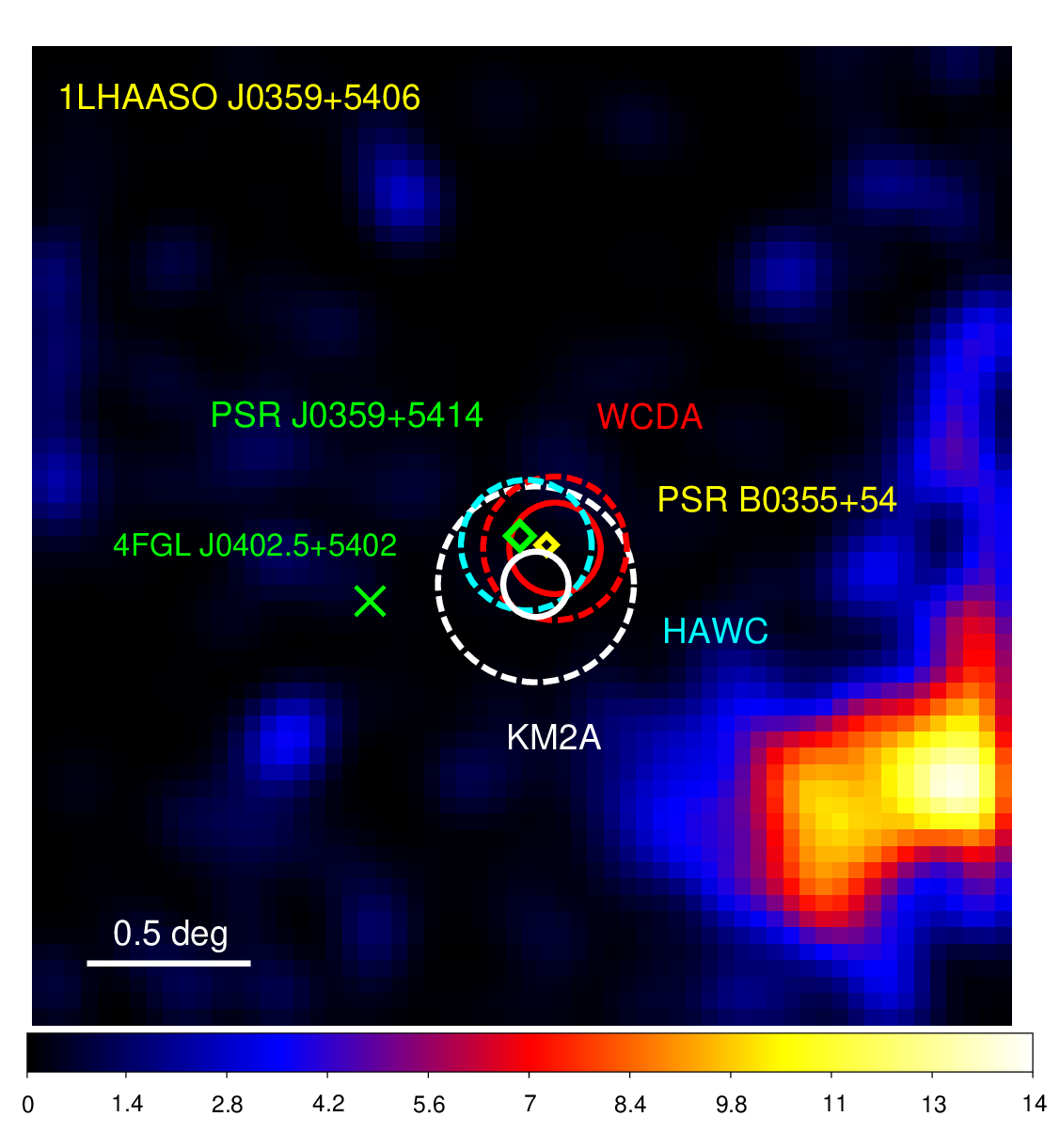}{0.33\textwidth}{(b)}
          \fig{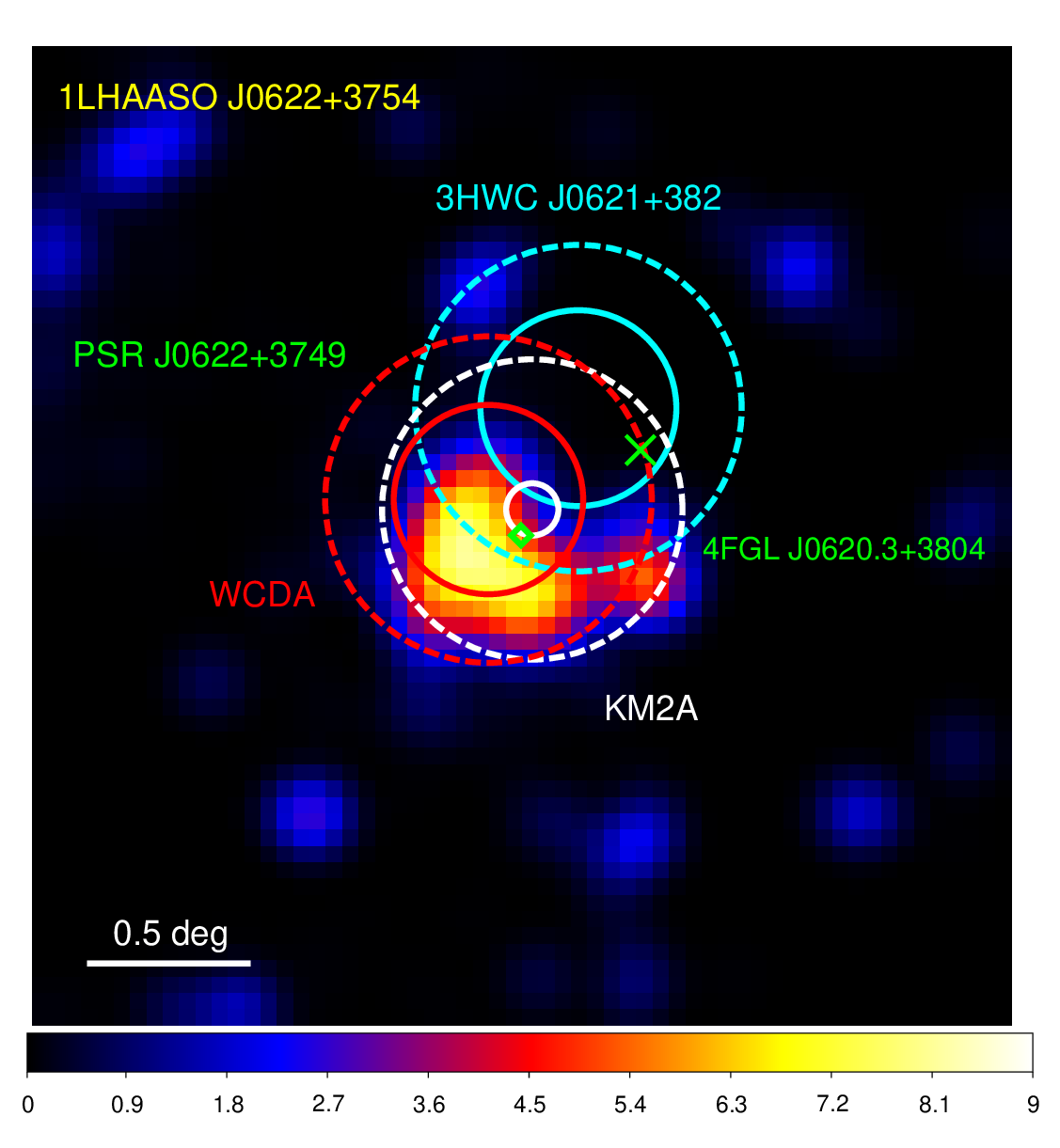}{0.33\textwidth}{(c)}
          }\vspace {-5mm} 
\gridline{\fig{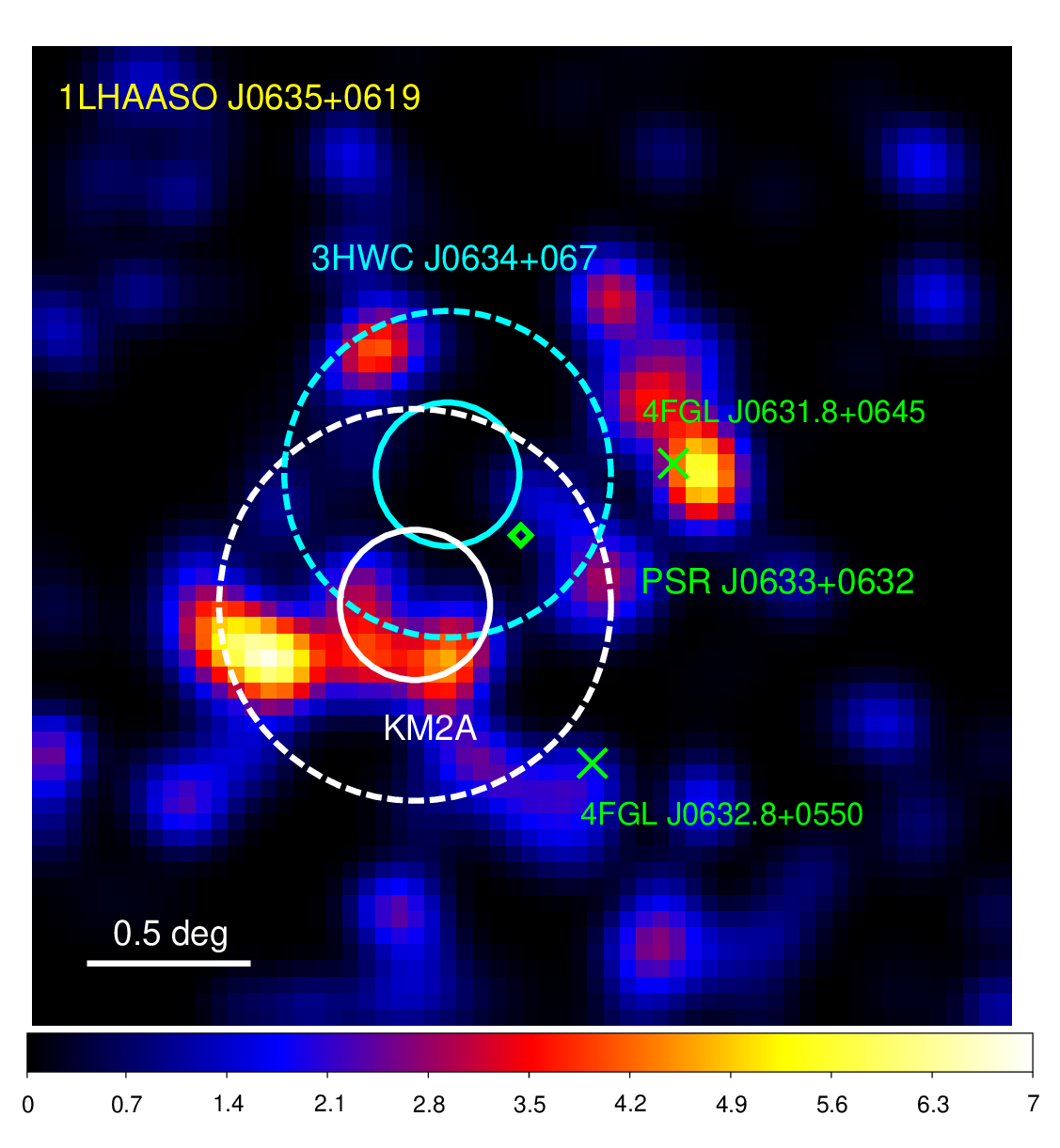}{0.33\textwidth}{(d)}
          \fig{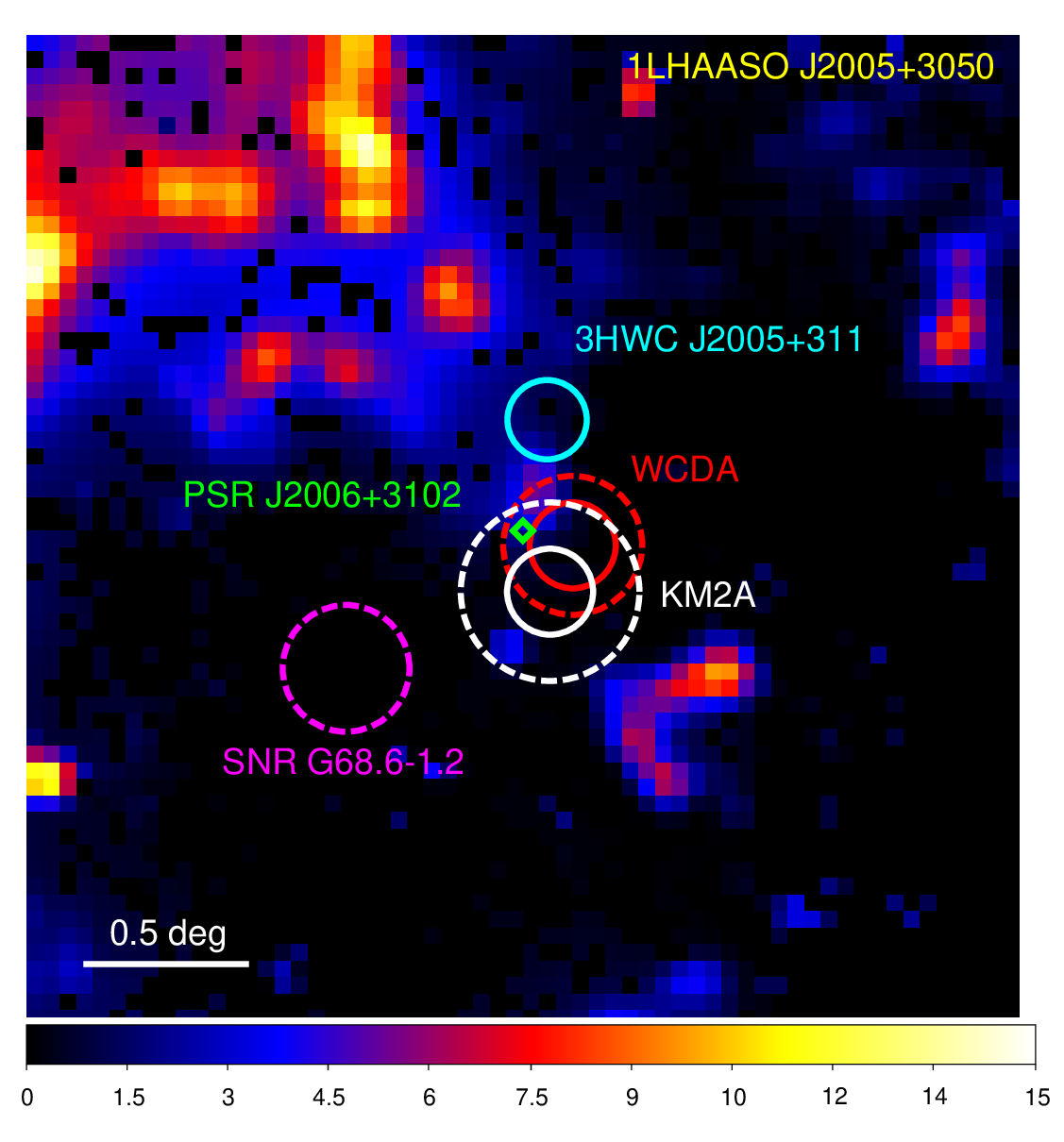}{0.33\textwidth}{(e)}
          \fig{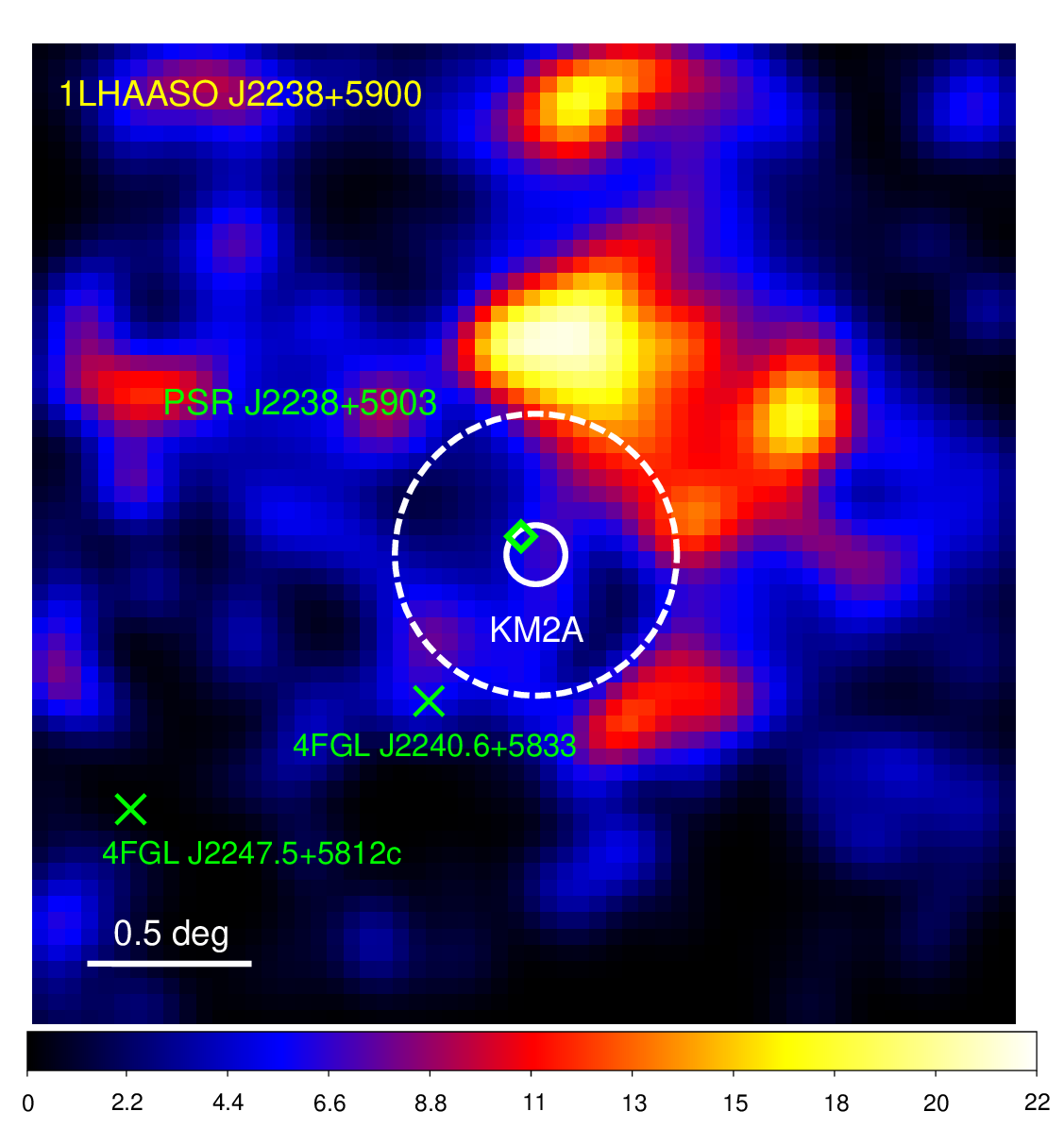}{0.33\textwidth}{(f)}
          }\vspace{-5mm} 
\caption{TS maps of the regions of the six pulsar targets in 
	0.1--500\,GeV, calculated from the off-pulse data of the pulsars. 
	Each panel has a size of $3^{\circ} \times 3^{\circ}$
centered at a pulsar target.
Green diamonds and crosses mark the positions of the pulsars and nearby 
	\fermi-LAT sources, respectively. The positional error circles
	and extension regions of LHAASO, HAWC, and HESS sources are marked
	by solid and dash circles respectively, and the name of the 
	corresponding 1LHAASO source is given at top-left of each panel. 
In the region of 1LHAASO J0359+5406,
	a non--\gr\ pulsar is indicated by a yellow diamond,  and in
	that of 1LHAASO J2005+3050, an SNR is shown as the magenta 
	dashed circle. 
\label{fig:tsmap}}
\end{figure*}

To show the non-detection results in the off-pulse data and provide a clear 
view of the source fields, we obtained 0.1--500\,GeV TS maps 
for the pulsars' regions. 
As indicated by the TS maps (Figure \ref{fig:tsmap}), 
no significant residual emissions are seen in any of the pulsar regions. 

\section{Pulsars and their associated VHE sources}
\label{sec:res}

Below, based on the analysis results we obtained for each pulsar target
(cf., Figures~\ref{fig:spec} \& \ref{fig:tsmap}), we briefly describe
the properties of the pulsars and their likely associated VHE sources 
in the following sections. We also searched for X-ray observational results
for the pulsars, which helps us learn about the properties of their PWNe. When 
needed, we analyzed the archival X-ray data by ourselves.

\subsection{PSR J0248$+$6021}
PSR J0248 is a $\gamma$-ray pulsar, with its radio pulsations first detected 
by the Nan\c{c}ay radio telescope \citep{frc+97}.  
At the pulsar's position, 
no X-ray emission was detected, and no evidence showed any
extended X-ray emission 
around the pulsar \citep{mdc11}. The unabsorbed X-ray flux upper limit 
was $9 \times 10^{-13}\,$erg\,cm$^{-2}$\,s$^{-1}$ in 
0.3--10.0 keV \citep{mdc11}. The distance used in this work was derived 
by \citet{tpc+11}. 

LHAASO detected an extended source, 1LHAASO J0249+6022, that is in positional 
coincidence with J0248.  The extension of the source is $\sim 0\fdg38$. 
In this region, no excess $\gamma$-ray 
emission was detected in off-pulse phase data analysis 
(see Figure \ref{fig:tsmap}). We noted that the region is rather clean, within
which no SNRs are listed in the SNR catalog 
SNRcat\footnote{\url{http://snrcat.physics.umanitoba.ca}}. 
Therefore we suggest that the extended TeV emission, 1LHAASO J0249$+$6022, 
is a TeV halo candidate powered by PSR J0248.  

\subsection{PSR J0359$+$5414}

The region of this pulsar is clean with no residual emissions detected in
the off-pulse data (Figure~\ref{fig:tsmap}). In X-rays, a weak PWN was 
detected with a luminosity
of $\simeq 2.8\times 10^{31}$\,erg\,s$^{-1}$ at a pseudo distance of
3.45\,kpc \citep{zks18}. Extended TeV emission at the region was reported 
by HAWC, with a size of $0\fdg2 \pm 0\fdg1$,
and this detection was matched by the LHAASO detection results.

The VHE source was already posited to be a TeV halo candidate powered 
by J0359 \citep{aaa+23}. 
To distinguish TeV halos and PWNe, it has been discussed that
the VHE $\gamma$-ray emissions are from a larger region than those of 
PWNe \citep{lab+17, loa+22, aaa+23}. However, this case is complicated by
the existence of a nearby radio pulsar B0355+54 (Figure~\ref{fig:tsmap}), 
which has 
a spin-down luminosity of $\dot{E} = 4.5 \times 10^{34}$\,erg\,s$^{-1}$
and a characteristic age of $\tau$ = 564\,kyr, and as such,
this radio pulsar's possible association with the VHE source
could not be excluded \citep{aaa+23}. 


\subsection{PSR J0622$+$3749}

This pulsar is radio quiet, with an X-ray flux upper limit of 
$1.4 \times 10^{-14}$\,erg\,cm$^{-2}$\,s$^{-1}$ in 0.1--2.0 keV \citep{pb15}. 
In the region, LHAASO detected extended VHE $\gamma$-ray emission named 
LHAASO J0621$+$3755, and it is likely a TeV halo \citep{aaa+21}. 
In the 1LHAASO catalog, 1LHAASO J0622$+$3754 was assigned to be associated 
with LHAASO J0621$+$3755, with the separation between them being
only $0\fdg03$. Our analysis of the off-pulse data verified the emptiness of
the field at GeV $\gamma$-rays.

The distance of the pulsar was estimated to be 1.6\,kpc by \citet{pga+12},
where the pulsar's \gr\ luminosity $L_{\gamma}$ in 0.1--100\,GeV 
was estimated from a $L_{\gamma}$-$\dot{E}$ relationship that was
derived based on the \gr\ pulsars with distance measures (for details 
see \citealt{sdz+10, pga+12}).
We re-estimated the distance by using the flux value given in the recent 
4FGL-DR4,
and found a value of 1.4 kpc. However this value can be highly different
from the actual one. Another method to estimate the distance is
to require $L_{\gamma}\leq \dot{E}$, which sets an upper limit of
3.47\,kpc for the distance, and if considering $L_{\gamma}\sim 0.1\dot{E}$, 
the distance
would be $\sim$1.1\,kpc. We adopted 1.1\,kpc for J0622 but with an
upper limit of 3.47\,kpc.

\subsection{PSR J0633$+$0632}

This pulsar is also radio quiet, first detected at $\gamma$-rays
by \fermi-LAT \citep{aaa+09}. In this source region, diffuse X-ray emission was 
detected and identified as a PWN \citep{rkp+11, dko+20}. For J0633 and its PWN,
the unabsorbed X-ray fluxes were 
$3.31^{+0.58}_{-0.62} \times 10^{-14}$\,erg\,cm$^{-2}$\,s$^{-1}$ and 
$1.17^{+0.11}_{-0.13} \times 10^{-13}$\,erg\,cm$^{-2}$\,s$^{-1}$ in 
2--10\,keV, respectively \citep{dko+20}. The source distance was 
discussed to be within a range of 0.7--2\,kpc,
based on an interstellar absorption-distance relationship \citep{dko+20}.

The LHAASO detection indicated that the VHE source has a hard emission, as
the WCDA observations only provided a flux upper limit \citep{1lhaaso}.
Our analysis of the off-pulse data provided a flux upper limit of
$\sim 10^{-13}$\,erg\,cm$^{-2}$\,s$^{-1}$ in the GeV energies.
Considering that the source 1LHAASO J0635$+$0619 (as well as 
3HWC~J0634+067) is a TeV halo candidate powered by PSR J0633, 
the VHE emission is likely from a larger region than that of the X-ray PWN. 
\citet{kbp+23} searched for X-ray counterparts of TeV halos (so-called 
X-ray halos) around 5 pulsars that include J0633. However, no such extended
emission was found.

\begin{figure*}[htbp]         
\gridline{\fig{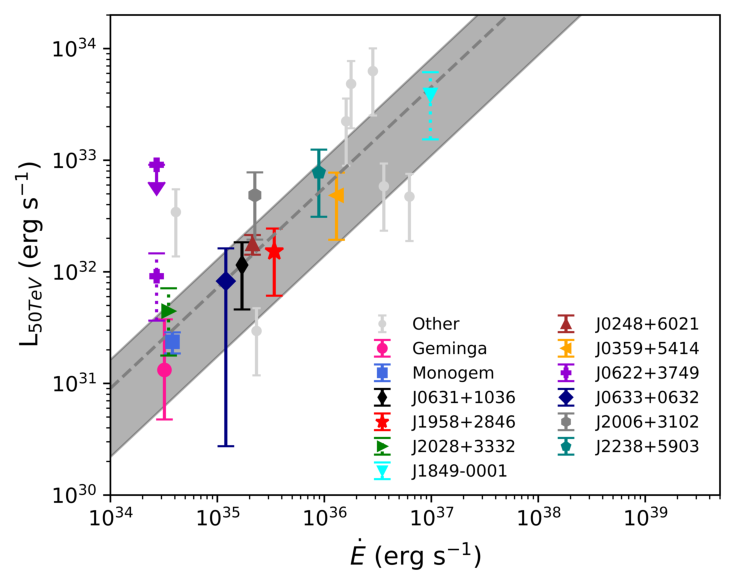}{0.33\textwidth}{(a)}
          \fig{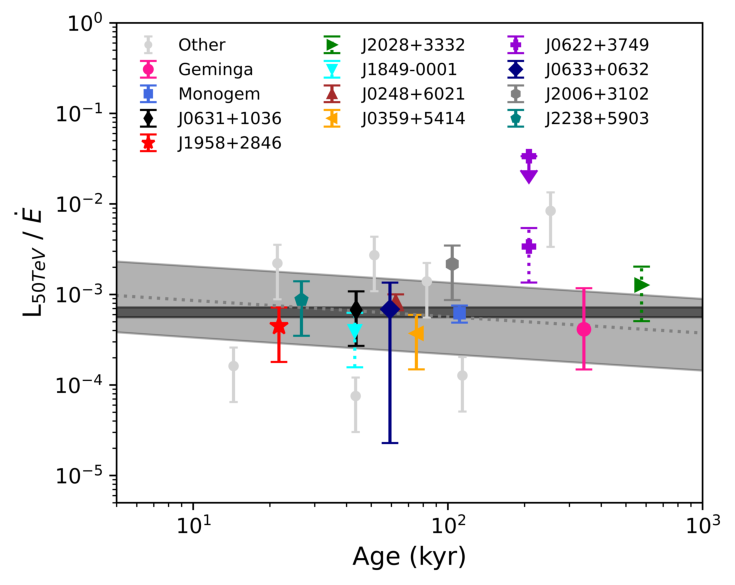}{0.33\textwidth}{(b)}
           \fig{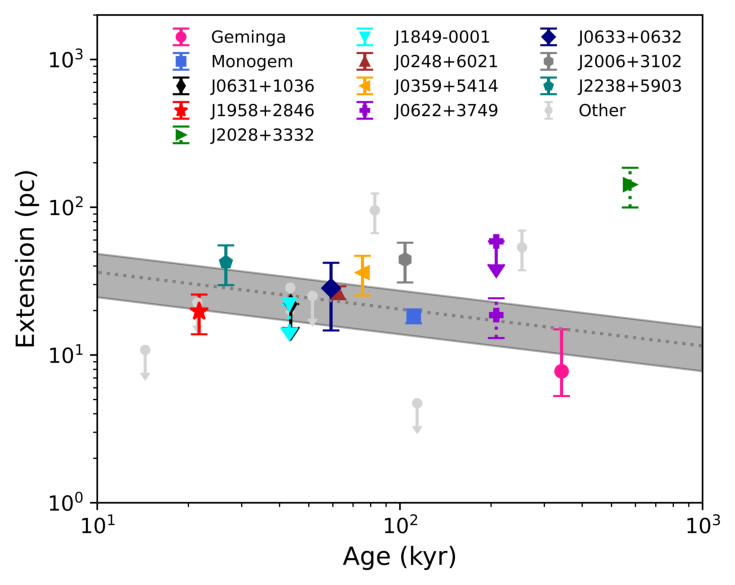}{0.33\textwidth}{(c)}}
	\caption{{\it Left:} relationship between luminosities 
	of (candidate) TeV halos at 50\,TeV $L_{\rm 50TeV}$ 
	(from the 1LHAASO measurements) and the corresponding pulsars' 
	spin-down energy $\dot{E}$, $L_{50 TeV} \propto \dot{E}^{0.9}$ 
	(dashed line). The shaded area indicates the 1$\sigma$ error 
	range.
	{\it Middle:} $L_{\rm 50TeV}/\dot{E}$
	being a function of the pulsars' characteristic ages $\tau$,
	$L_{\rm 50TeV}/\dot{E}\sim 1.3\times 10^{-3}\tau_{\rm kyr}^{-0.18}$
	(dotted line with the shaded area indicating the 1$\sigma$ 
	error range),
	or being $\sim 6.4 \times 10^{-4}$ (dark line region, with
	the width indicating the 1$\sigma$ error range).
	{\it Right:} physical sizes $S$ of (candidate) TeV halos
	as a function of $\tau$, $S\sim \tau_{\rm kyr}^{-0.25}$ (dotted line
	with the shaded area indicating the 1$\sigma$ error range).
	For details, see Section~\ref{sec:dis}.
\label{fig:prop}}
\end{figure*}  

\subsection{PSR J2006$+$3102}

This radio pulsar was reported with a distance of 4.7\,kpc  
in \citet{nab+13}, but the updated value is 6.035\,kpc in the Australia Telescope National Facility (ATNF) pulsar catalog \citep{man+05}.
Very limited information is available for this pulsar.
We searched the {\it Chandra} and {\it XMM-Newton} archival data, but
no observations were found. Using a set of data 
(Obsid : 03103085001, exposure time = 1.1\,ks) obtained with the X-ray 
Telescope (XRT) onboard {\it the Neil Gehrels Swift Observatory (Swift)}, we
derived a 3$\sigma$ upper limit of 0.01\,cts\,s$^{-1}$ in 0.3--10\,keV
at the pulsar's position.
The corresponding energy-flux upper limit was
$9.0\times 10^{-13}$\,erg\,cm$^{-2}$\,s$^{-1}$, where we assumed
a PL source spectrum with an index of 2 and hydrogen column density 
$N_{\rm H} = 8.27 \times 10^{21}$\,cm$^{-2}$ (towards the 
source direction, from \citealt{hi4pi+16}).

Close to the edge of the extension region given by the LHAASO KM2A, 
there is an SNR, 
G68.6$-$1.2 (Figure~\ref{fig:tsmap}), which, however, is faint and poorly
defined according to the SNRcat. Given its poorly known properties and
relatively large separation ($\sim 0\fdg68$) from the VHE source, 
it is not clear if the SNR can be connected to the latter.
We noted that 3HWC J2005$+$311 is also located in this region \citep{3hwc},
and its spectrum is similar to that of 1LHAASO J2005$+$3050.
However, the positions of the two sources do not overlap. The relation 
between them remains to be resolved from further observational results.

\subsection{J2238$+$5903}

J2238 also has very limited information available. An X-ray flux upper 
limit was reported by \citet{pb15} to be 
$4.4 \times 10^{-14}$\,erg\,cm$^{-2}$\,s$^{-1}$ in 0.1--2 keV 
(where a PL with index = 1.7 was assumed).
The LHAASO WCDA observations were influenced by the Galactic Diffuse Emission 
\citep{1lhaaso}, and we did not consider the WCDA measurements.

\section{Discussion}
\label{sec:dis}

Following our previous work on identifying candidate pulsar TeV halos 
\citep{zwx23,zw23} by mainly
analyzing the off-pulse GeV data of \gr\ pulsars in the fields of VHE sources,
from which any residual emissions may help reveal their nature as possibly
being
primary Galactic sources, such as SNRs or PWNe, we further found six candidates
because of the non-detection of any significant residual emissions.
The pulsars' properties, including information for their X-ray emissions,
are summarized in Table~\ref{tab:psr}. 
As discussed in \citet{zw23}, there may contain a relationship between
the TeV halos' luminosity at 50\,TeV, $L_{\rm 50TeV}$, 
and corresponding pulsars' spin-down energy $\dot{E}$. 
This relationship helps indicate the fraction
of the total energy spent on powering the TeV halos. 
Since most of the sources (including those presented in the Appendix)
in this work have been detected by LHAASO KM2A in 25--100\,TeV,
we thus also estimated their
$L_{\rm 50TeV}$ from the differential fluxes at 50\,TeV given in the
LHAASO results \citep{1lhaaso}. The $L_{\rm 50TeV}$ values are
provided in Table~\ref{tab:psr}.
Fitting the data points that include four sources in \citet{zw23}
and five sources in this work (excluding J0622 whose distance is highly
uncertain), we obtained 
$L_{\rm 50 TeV} = 2.27^{+1.82}_{-1.72} \dot{E}^{0.90^{+0.02} _{-0.01}}$,
with a reduced $\chi^2$ value of $\simeq$0.8 for 7 degree of freedom
(DoF), where we assumed a 30\% uncertainty for distances (this uncertainty
was dominant). 
We used the Markov Chain Monte Carlo (MCMC) code 
{\tt emcee} \citep{fhl+13} for the fitting,
since it conveniently provides error ranges.
It can be noted that the $L_{\rm 50 TeV} \sim \dot{E}^{0.9}$ relationship
(see Figure~\ref{fig:prop}), reported in \citet{zw23}, still holds.

Another relationship we tested was $L_{\rm 50 TeV} / \dot{E}$ being
either a function of 
the pulsars' characteristic ages $\tau$ or a constant. Fitting 
the data points, we obtained 
$L_{\rm 50 TeV} / \dot{E}$ = $1.3^{+1.8}_{-0.8} \times 10^{-3} \tau_{\rm kyr}^{-0.18^{+0.23}_{-0.21}} $ (where $\tau$ is in units of kyr) with 
reduced $\chi^2 \simeq$0.8 for DoF=7, or
$L_{\rm 50 TeV} / \dot{E} = 6.4 \pm 0.8 \times 10^{-4}$ with
reduced $\chi^2\simeq 0.6$ for DoF=8. Both results
are also very similar
to those previously obtained in \citet{zw23}. For the first result,
the large uncertainty for the $\tau$ index indicates its value close to zero,
and thus the second result,
$L_{\rm 50 TeV} / \dot{E}$ being a constant (as in \citealt{zw23}), is 
preferred.

In addition, we also tested the physical sizes $S$ of the VHE sources as a 
function of $\tau$. The sizes were derived from the extension sizes in degrees,
as summarized in Table~\ref{tab:psr} from the LHAASO KM2A measurements.
We obtained
$S = 64.51^{+21.54}_{-21.04}\times \tau_{kyr}^{-0.25^{+0.09}_{-0.07}}$\,pc, 
with reduced $\chi^2 \simeq 2.1$ for DoF=6.
The uncertainties are large, and there is a source, J2028+3332 \citep{zw23},
significantly deviating from the relationship (although the source's distance
is uncertain). In any case, there is a possible
older-and-smaller trend, which could be an interesting feature that may
reveal the evolutionary processes of electron/positron ejection of pulsars 
and halos. Further observational results obtained from more collected data 
with LHAASO may verify this trend.

As we also searched for other potential TeV halo candidates from among mainly
1LHAASO sources, there are seven of them whose properties may provide
hints as to their possible TeV-halo nature
based on different studies (See Appendix Section~\ref{sec:src}). We showed
their corresponding properties (Table~\ref{tab:psr}) in 
Figure~\ref{fig:prop}. As can be seen, they generally have large scattering
around the relationships we obtained above. In particular, five of them were
compact sources (see Appendix Section~\ref{sec:src} and 
Figure~\ref{fig:tsmapa}) in the LHAASO KM2A measurements. Because different
VHE observational facilities have different sensitive energy bands and
spatial resolutions, we did not try to replace the KM2A results with
those from other facilities.  Thus, most of these sources currently do not 
fit in the $S \sim \tau$ relationship at all.
From this comparison, we may conclude that either they are not TeV halos
or their emissions may contain significant contributions from other sources,
which would be in agreement with the various results from many multi-energy
studies about them (Appendix Section~\ref{sec:src}).

\begin{figure}[htbp]      
	\includegraphics[width=0.48\textwidth]{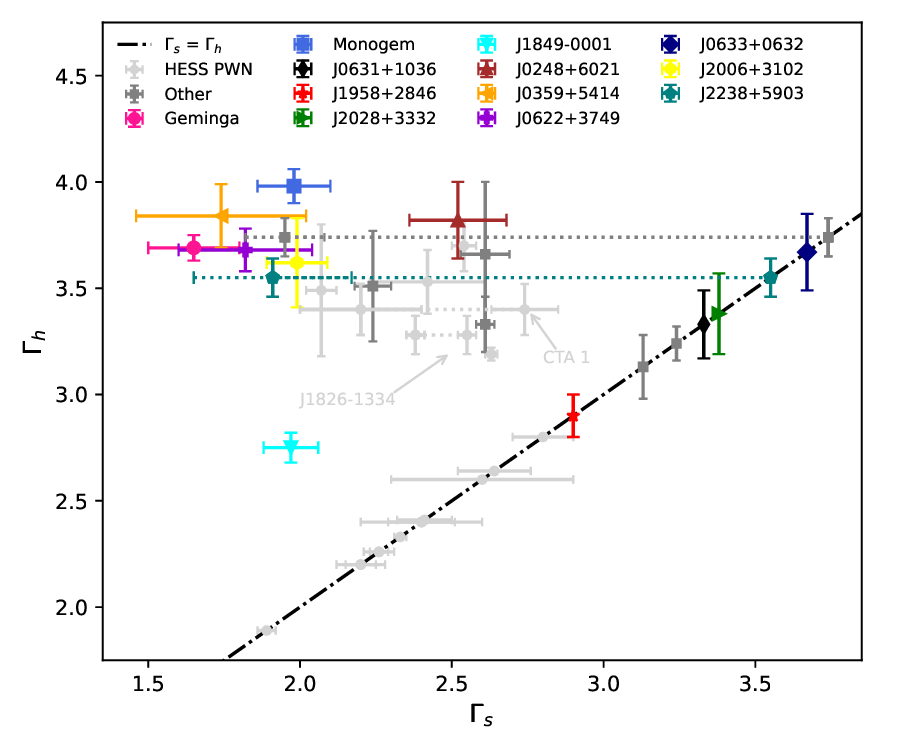}
	\caption{Power law indices of the (candidate) TeV halos and 
	HESS-confirmed PWNe. Values of the first group
	are from
	LHAASO WCDA ($\Gamma_s$ in 1--25\,TeV) and KM2A ($\Gamma_h$ in
	25--100\,TeV), and those of the latter are mostly from 
	HESS in 1--10\,TeV \citep{haa+18}. When there is only one 
	measurement, either $\Gamma_s$ or $\Gamma_h$,
	the source is put at the $\Gamma_h = \Gamma_s$ line (dash-dotted).
	Because two WCDA measurements suffered GDE, the sources
	are also put at the $\Gamma_h = \Gamma_s$ line as well, indicated by
	the dotted lines. HESS~1825$-$137 (or 
	1LHAASO~J1825$-$1337u, associated with PSR J1826$-$1334) and
	CTA~1 (\citealt{ali+13}; or 1LHAASO~J0007+7303u) have different reported
	$\Gamma_s$ values, and both values of each of them are shown
	(connected with a grey line and pointed with an arrow). The PWN
	of the Vela pulsar is the lowest grey data point along the 
	$\Gamma_h = \Gamma_s$ line (with $\Gamma_s < 2$).
\label{fig:index}}
\end{figure} 

Finally, \citet{haa+18} studied all PWNe and candidates in 1--10\,TeV. 
On the basis of their results, one conclusion that may be drawn is PWNe
tend to have a soft spectrum with the PL index $\Gamma_s>2$. By comparison,
as pointed
by \citet{zw23}, candidate TeV halos often show hard spectra with $\Gamma_s<2$.
We further explored this possible feature by constructing 
Figure~\ref{fig:index}, in which the PL indices of the candidate TeV halos
(as well as the sources described in the Appendix) and the HESS-confirmed PWNe
are shown, where the hard PL indices $\Gamma_h$ are from the LHAASO KM2A 
25--100\,TeV
measurements. Some of the sources, in particular those HESS PWNe, were only 
detected in one energy band (such as the soft 1--10\,TeV band), and we put
these sources at the $\Gamma_h = \Gamma_s$ line; note that the
error bars indicate the measurements at which energy band are known.
It is clear
to see that most candidate TeV halos either show emissions with $\Gamma_s<2$, 
or simply have detectable hard TeV emissions (only with known 
$\Gamma_h$ in 25--100\,TeV). 
By comparison, PWNe have soft emissions with $\Gamma_s>2$ or do not
have any detectable hard TeV emissions (those at the $\Gamma_h = \Gamma_s$ 
line).
One notable source of the PWNe is that of the Vela pulsar, which has 
$\Gamma_s<2$ (the data point
at the low left along the $\Gamma_h = \Gamma_s$ line in Figure~\ref{fig:index}).
On the other hand, one exception among the candidate TeV halos is
1LHAASO J0249+6022 (associated with PSR J0248), which has $\Gamma_s>2$. Detailed
studies of this source may help understand the cause of the deviation.
In any case, the comparison strengthens our previous suggestion in
\citet{zw23} that TeV halos are different from PWNe by having hard emissions.

\begin{acknowledgements}
	We thank the anonymous referee for helpful comments.
	This research is supported by the Basic Research Program of Yunnan 
	Province (No. 202201AS070005), the National Natural Science Foundation 
	of China (12273033), and the Original
Innovation Program of the Chinese Academy of Sciences (E085021002).
D.Z. acknowledges the support of the science research program for graduate 
	students of Yunnan University (KC-23234629).
\end{acknowledgements}

\bibliography{lhaasohalo.bib}{}
\bibliographystyle{aasjournal}



\appendix

\restartappendixnumbering

\section{Timing analysis and brief introduction for seven candidates}
\label{sec:src}

\begin{figure*}
\gridline{\fig{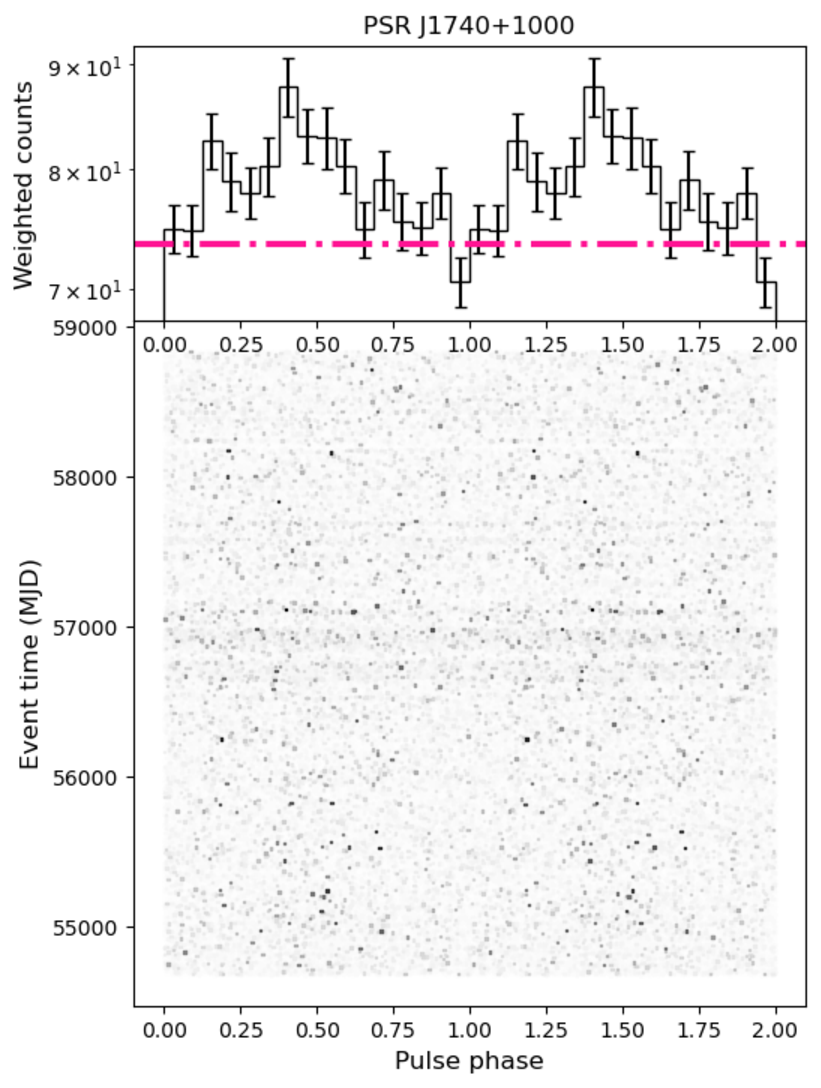}{0.3\textwidth}{}\hspace{-15mm}
          \fig{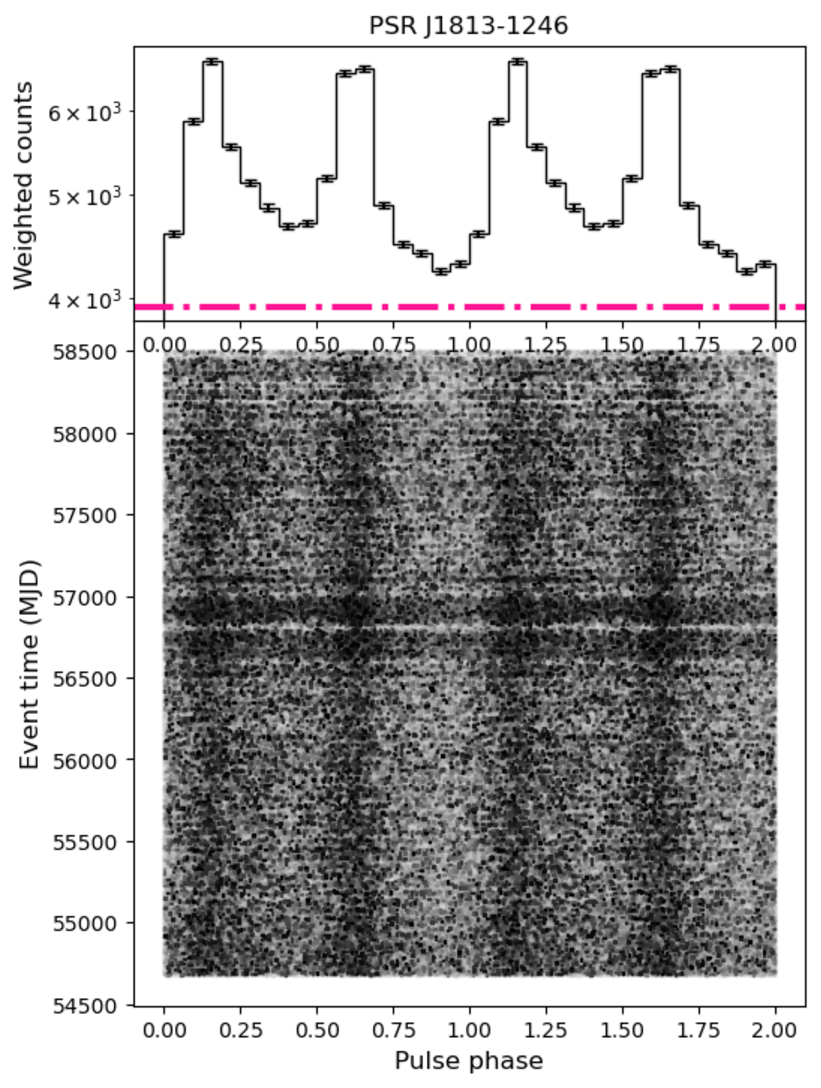}{0.3\textwidth}{}\hspace{-15mm}
	  \fig{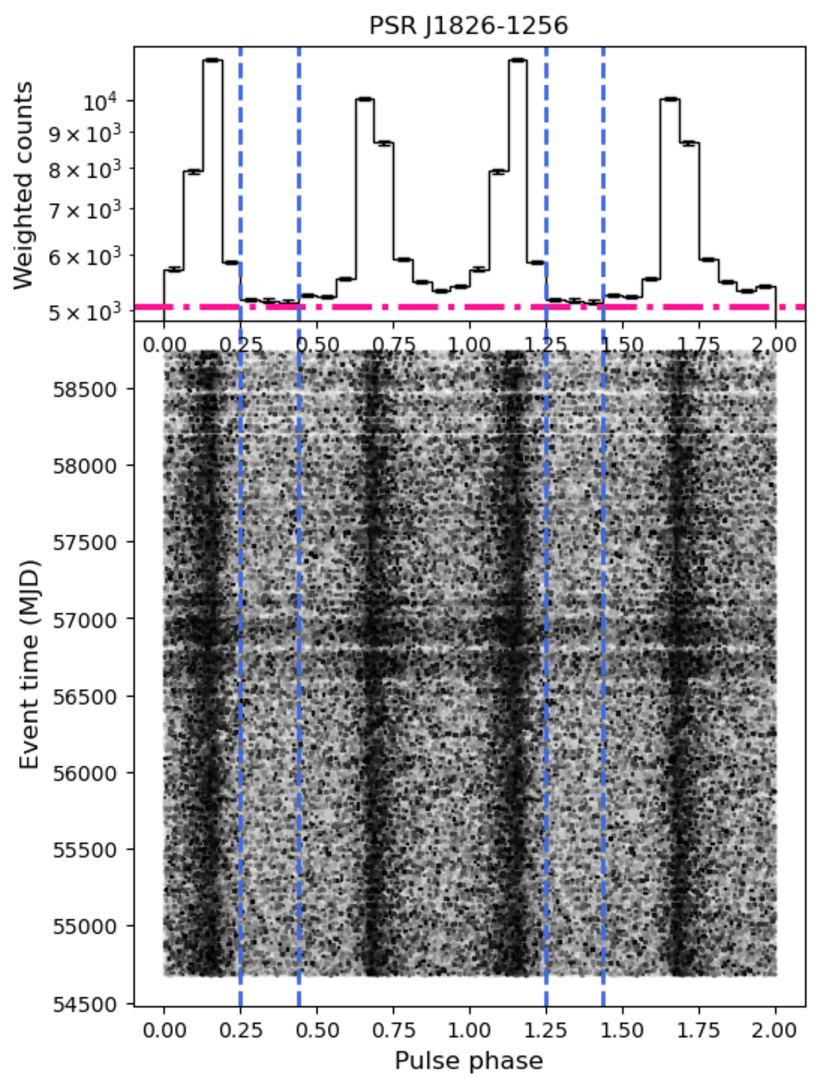}{0.3\textwidth}{}}
\caption{Pulse profiles (top) and two-dimensional phaseograms (bottom) 
	of three \gr\ pulsars. Only for PSR J1826$-$1256, an off-pulse phase
	range can be defined, which are marked by dashed lines.
\label{fig:phasea}}
\end{figure*}

We also collected information for seven additional VHE sources, given their
particular features revealed from observational and theoretical studies.
Three of the corresponding pulsars have \gr\ emissions, but two of them
(J1740+1000 and J1813$-$1246) do not have clear off-pulse phase ranges.
The timing solutions we used are given in Table~\ref{tab:timinga}, and the
pulse profiles are shown in Figure~\ref{fig:phasea}. The other four pulsars
do not have detectable \gr\ emissions in \fermi-LAT data. We analyzed
the \fermi-LAT data using the same processes described in Section~\ref{sec:da},
and calculated the TS maps for each of the sources. For the field of 
PSR~J1826$-$1256, the TS map was from the pulsar's off-pulse data, and for
all the other sources, the TS maps were from the whole data but with
all known sources given in 4FGL-DR4 removed.
As the information
in the following sections indicates, most of the sources are in complex
regions (see Figure~\ref{fig:tsmapa}) and their nature is still 
under different investigations.

\begin{deluxetable*}{cccccc}
\tablecaption{Timing solutions for three pulsars
\label{tab:timinga}}
\tablewidth{0pt}
\tablehead{
\colhead{Pulsar} & \colhead{End time} & \colhead{$f$} & \colhead{$f_1$/$10^{-12}$} &
\colhead{On-pulse} & \colhead{Off-pulse}  \\
\colhead{}            & \colhead{(MJD)} & \colhead{(Hz)} & \colhead{(Hz s$^{-1}$
)} & \colhead{} & \colhead{}
}
\startdata
                     J1740$+$1000 & 58839&6.489469647 & $-$0.89830   & --        &
--         \\
                     J1813$-$1246      & 58500&20.80158238 & $-$7.59705   & --
 & --         \\
                     J1826$-$1256      & 58738&9.071227968 & $-$9.97369   & 0--0.25, 0.4375--1        & 0.25--0.4375         \\
\enddata
\tablecomments{Frequencies are obtained from 3PC}
\end{deluxetable*}

\subsection{PSR B0540$+$23}
This pulsar was first detected by Jodrell Bank Mark IA radio telescope
\citep{dls72}. Analyzing the {\it XMM-Newton} archival data,
\citet{pb15} reported a faint X-ray point source
associated with the radio position.
No X-ray diffuse emission was detected by the extended ROentgen Survey 
with an Imaging Telescope Array \citep{kbp+23}.
HAWC detected $\sim 0\fdg5$ extended TeV emission, named HAWC J0543$+$233,
around the pulsar, which was suggested to be a potential TeV 
halo \citep{rfs17}. In 3HWC, the source was resolved
to two sources, 3HWC J0540$+$228 and J0543$+$231 \citep{3hwc}.
LHAASO detected a source, 1LHAASO J0542$+$2311u, positionally coincident
with 3HWC J0543$+$231 in the energy range of from 25 TeV to above 100 TeV,
and the detection showed a large extension of $\sim 0\fdg98$. The pulsar is
away from the reported positions of 1LHAASO J0542$+$2311u and 3HWC J0543$+$231
by $\sim 0\fdg29$ and $\sim 0\fdg37$, respectively.
We analyzed the \fermi-LAT data, but no GeV emissions were detected at
the pulsar's position. In the extension region given by LHAASO,
only one \fermi-LAT source, 4FGL J0544.4+2238, is known but at the region's
edge. This GeV source is an unidentified source with soft PL emission
(photon index was 3.33$\pm$0.17). It is rather hard to make connections between
the soft emission with the VHE sources. Thus in the rather clean region,
we considered the VHE emission possibly originates from a candidate TeV halo.

\subsection{PSR J1740$+$1000}
The Arecibo Telescope observations discovered this young pulsar, but it is
located away from the Galactic Plane \citep{mac+02}. An X-ray tail 
structure was found behind
the pulsar \citep{kmp+08}. VERITAS searched for TeV $\gamma$-ray emission
from the tail, but no emissions were detected \citep{bbb+21}. HWAC detected
a source in 14.8--274.0\,TeV in the region, which is likely also the LHAASO
source reportedly detected in the energy range from 25\,TeV to above 100\,TeV.
The pulsar is listed as a \fermi-LAT source but with a detection significance
of only $6.7\sigma$, and its pulsed \gr\ emission is likely too faint to be
clearly
identified (Figure~\ref{fig:phasea}). We tested to remove the GeV source 
and checked if there were
any residual emissions (the \fermi-LAT data analyzed were in 0.1--500 GeV
during the time period of from 2008-08-04 15:43:36 (UTC) to
2023-02-16 00:00:00 (UTC)). No such emission was found.
Given the non-detection of the PWN tail in VHE energies, we considered
the HAWC/LHASSO VHE source is a candidate TeV halo possibly
powered by the pulsar. It should be noted that if this scenario is the case,
it suggests that TeV halos could emit ultra-high-energy (UHE, $>100$\,TeV)
$\gamma$-rays, since the source had a TS value of 37.2 above 100 TeV in 
the LHAASO's detection.

\subsection{PSR J1809$-$1917}

The region is rather complex because of different VHE emission detections
plus additional sources revealed by related multi-energy studies.
A VHE source, HESS~J1809$-$193, was first reported in the region
and was considered as the emission originating from a candidate PWN 
associated with the pulsar \citep{aab+07}. Indeed, the PWN-like extended 
emission and its variations were observed in
X-rays \citep{kp07,abe+10,kkp+18,kyh+20}. In addition, there are at least
three SNRs, G11.0$-$0.0, G11.1$+$0.1, and G11.4$-$0.1, and several 
molecular clouds (MCs) located in this region \citep{cgp16}. The interaction 
between
G11.0$-$0.0 and a nearby MC was proposed to be the process
producing the VHE emission \citep{cgp16}. 

HGPS reported the source with a size of $\sim 0\fdg4$ \citep{hgps}, but
\citet{haa+23} re-analyzed the data and revealed 2 components,
an extended plus a compact one. While the compact component could be
either the VHE emission of the PWN or due to the SNR-MC interaction, the 
extended component was suggested to be a halo around the PWN of the pulsar
\citep{haa+23}. Correspondingly, HAWC and LHAASO each detected a source
in the region, while detailed properties, such as the extension size and 
flux, are different. 

Given the complexness of the region, any clear identification for 
the connections between different sources at multi-energies is not 
straightforward.  In any case, given the work in \citet{haa+23}, we considered
the extended component (matches the LHAASO WCDA detection) as being a possible
TeV halo.

\subsection{PSR J1813$-$1246}
The pulsar is radio-quiet, discovered by \fermi-LAT \citep{aaa+09}.
No obvious off-pulse phase range could be
determined in the GeV energy band (Figure~\ref{fig:phasea}), 
and no PWN was found in X-ray
observations \citep{mhp+14}. VHE emissions were detected by HESS, HAWC,
and LHAASO \citep{hgps, 3hwc, 1lhaaso} with a high positional coincidence. 
In \citet{hgps}, the VHE source was suggested to be associated with a relic 
PWN that is only detected in the TeV band. With limited information for 
the sources and the region, if we consider a relic PWN, an extended halo
forming from escaping particles from the PWN could be a possible scenario.
We thus included this source in the studies.

\subsection{PSR J1826$-$1256}
This pulsar is radio-quite. In this region, largely extended TeV emission,
named HESS~J1825$-$137 (see Section~\ref{sec:18}), was 
first detected by HESS \citep{aab+05}. In the HGPS, a new source, 
HESS J1826$-$130, was revealed at the northern edge of HESS~J1825$-$137 
\citep{hgps}, associated with PSR J1826$-$1256. Follow-up observations and 
studies identified this new source as the PWN, and it is a PeVatron 
candidate \citep{dgc+19,haa+20,bmg+22}. In the LHAASO results, 
1LHAASO J1825$-$1256u is positionally coincident with HESS J1826$-$130, but
the WCDA measurements provided higher fluxes and a larger size than
those given by HESS (e.g., the extension sizes were $\sim 0\fdg24$ and 
$\sim 0\fdg15$, respectively).

We were able to define an off-pulse phase range for the \gr\ emission
of the pulsar (Figure~\ref{fig:phasea}), but
no significant residual emission was detected in the off-pulse data.
In 3PC, the pulsar's distance was estimated to be 1.55 kpc. However,
\citet{kzs19} used interstellar reddening relationships towards the source 
direction and estimated a value of $\sim$3.5 kpc.
We included this pulsar in our studies, although it should be noted
that its characteristic age is rather young, 14.4\,kyr, and its PWN was
clearly detected in X-rays \citep{kzs19}. 

\begin{figure*}
\gridline{\fig{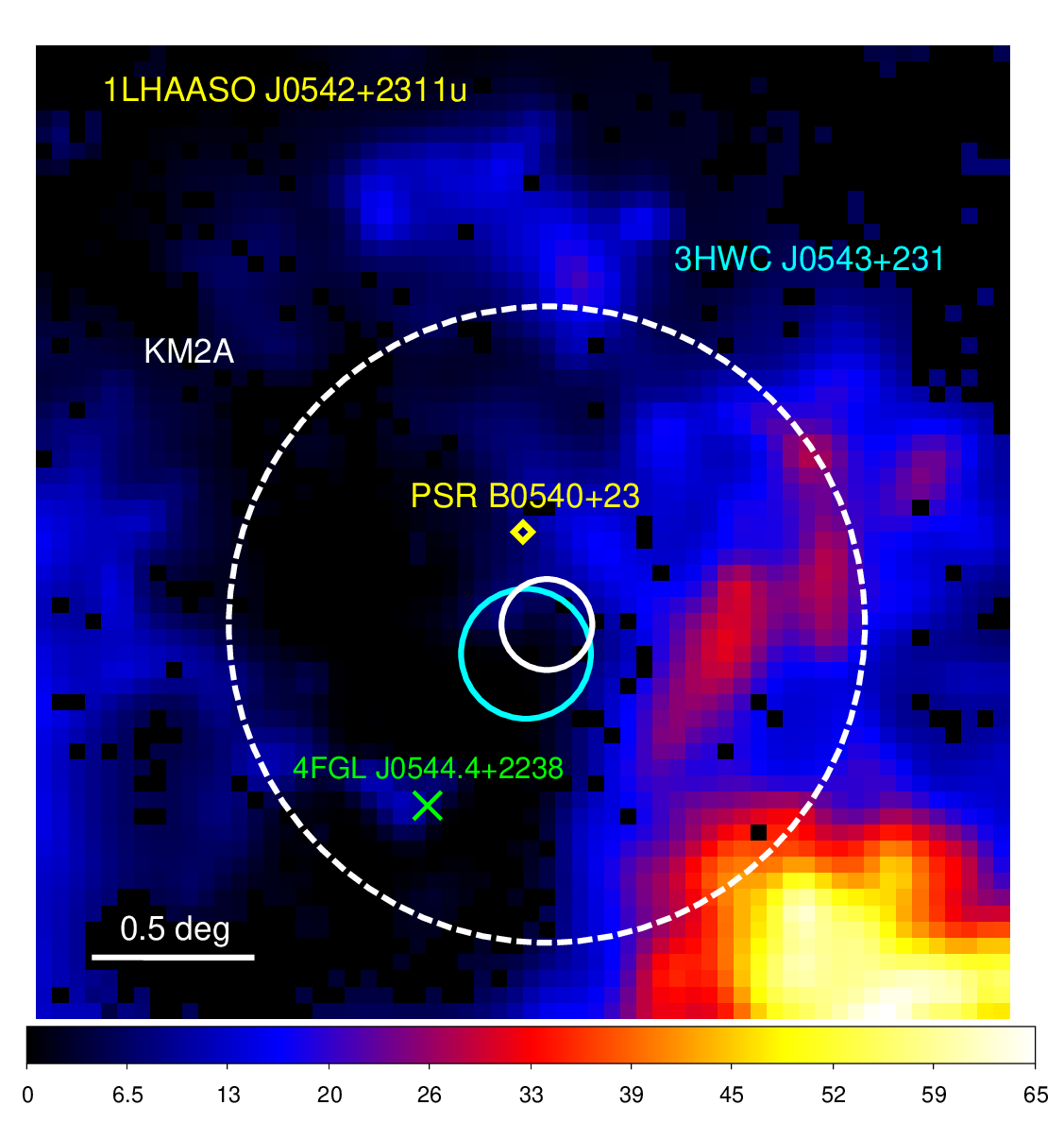}{0.33\textwidth}{(a)}
          \fig{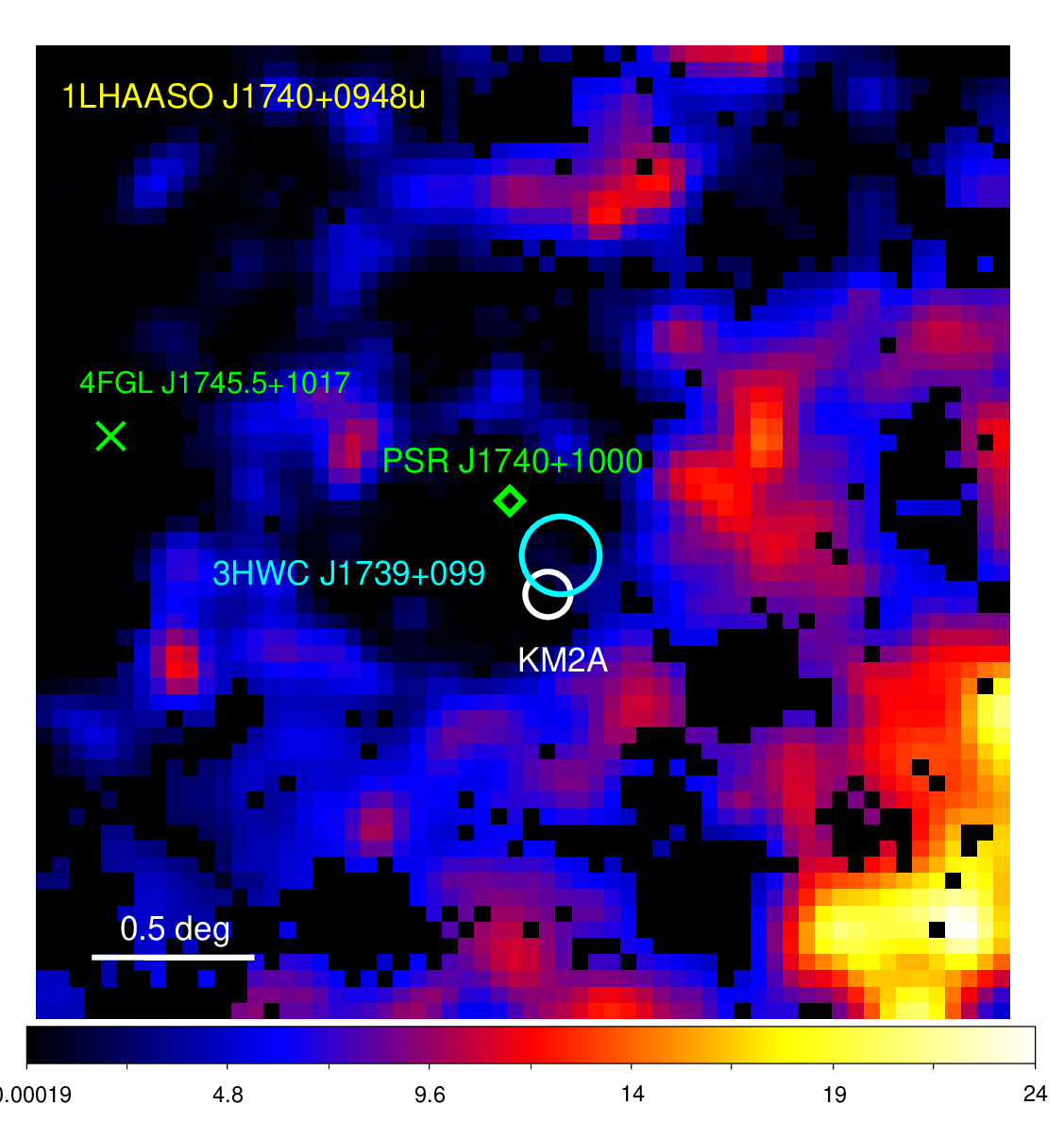}{0.33\textwidth}{(b)}
          \fig{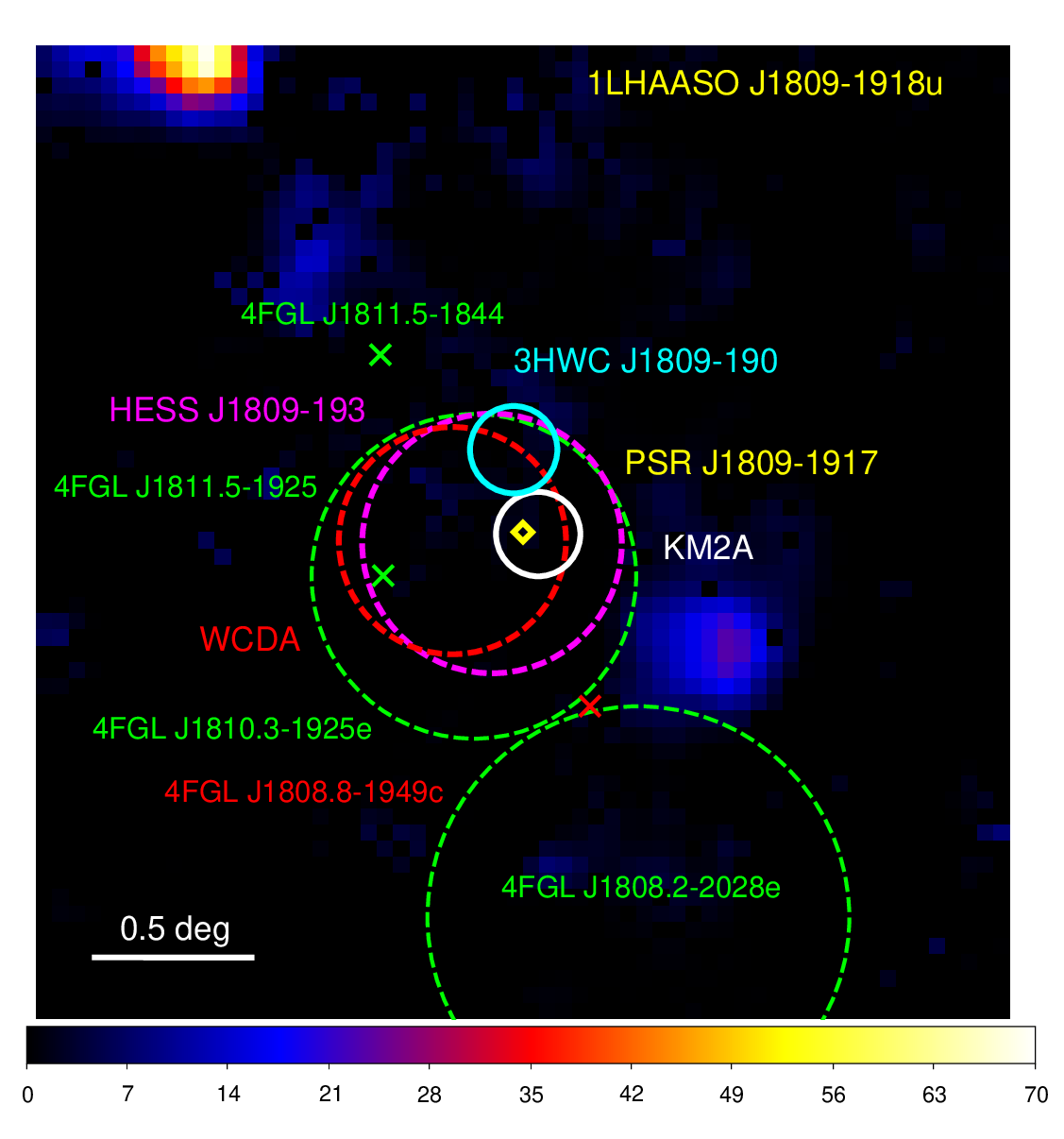}{0.33\textwidth}{(c)}
          }\vspace {-5mm}
\gridline{\fig{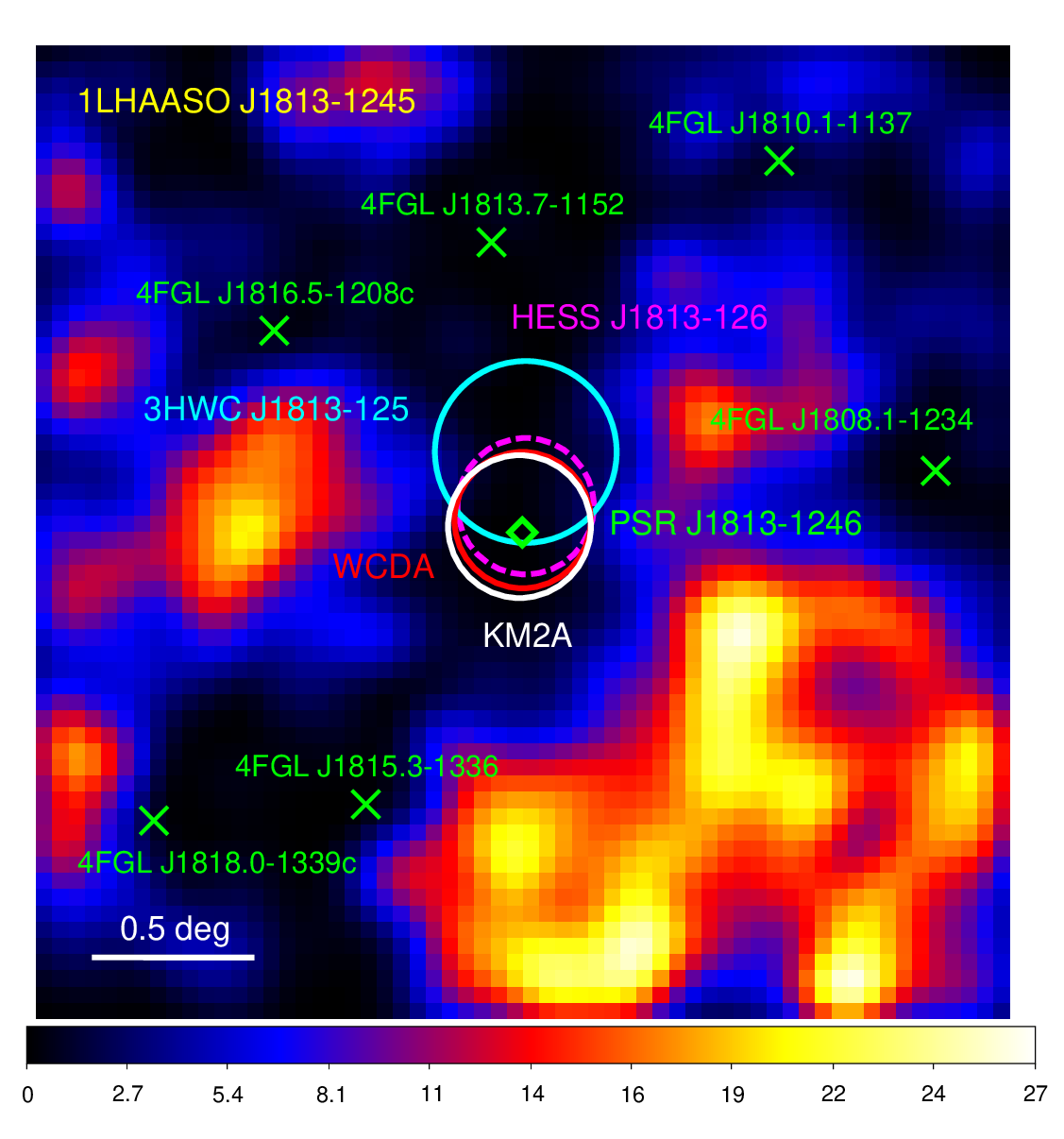}{0.33\textwidth}{(d)}
          \fig{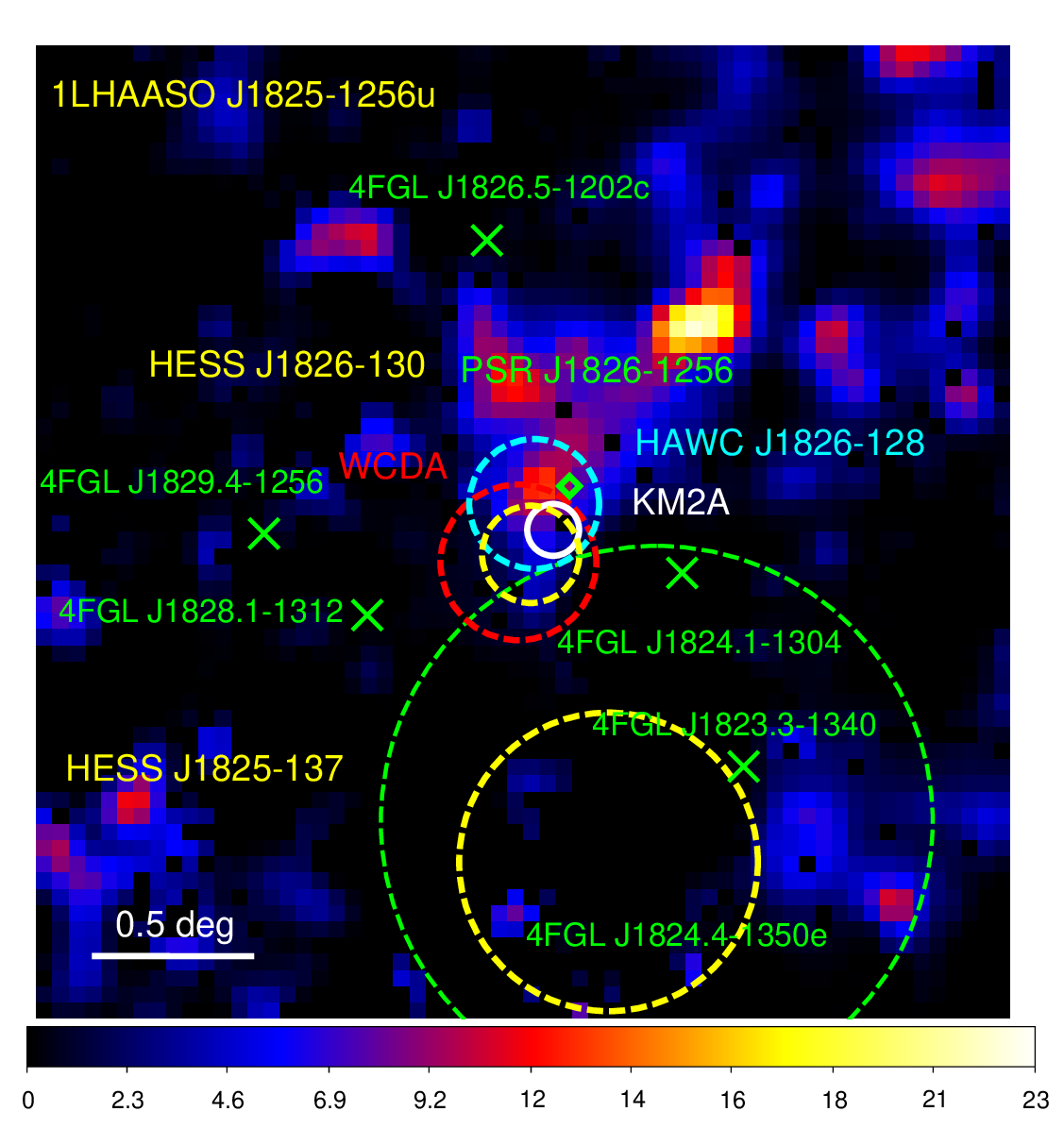}{0.33\textwidth}{(e)}
          \fig{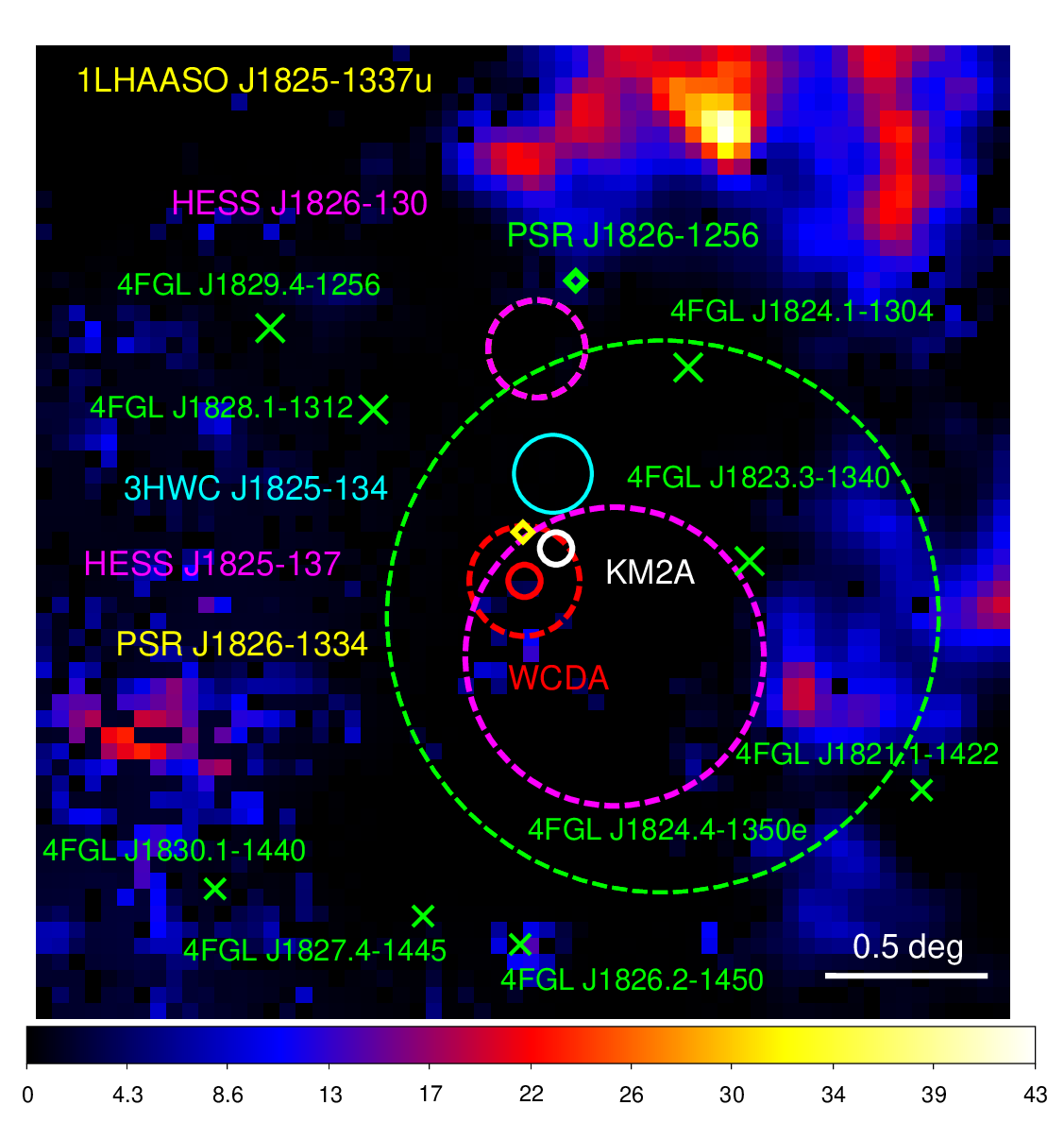}{0.33\textwidth}{(f)}
          }\vspace{-5mm}
\gridline{
             \fig{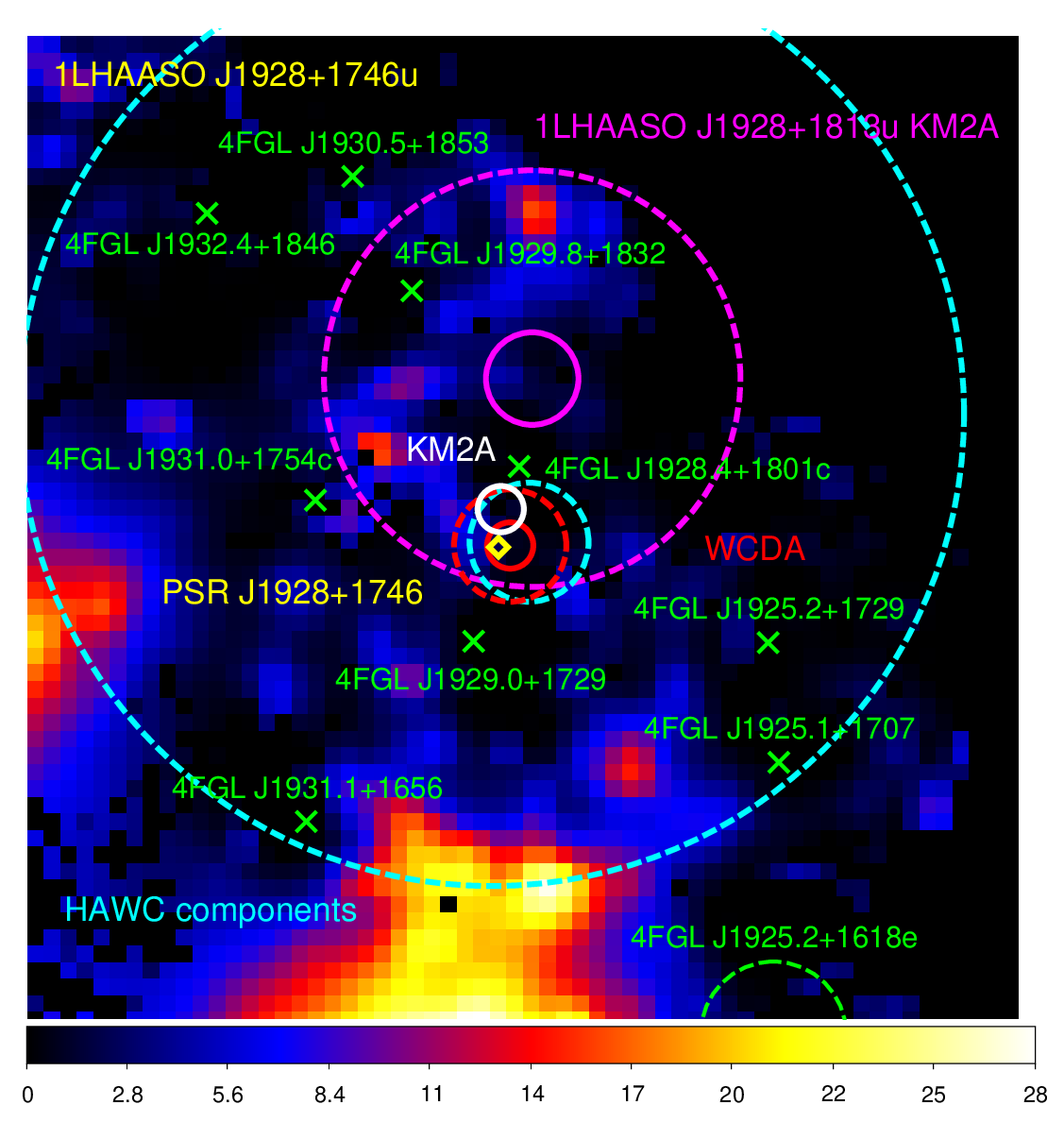}{0.33\textwidth}{(g)}}
\caption{TS maps of a size of $3^{\circ} \times 3^{\circ}$ in 0.1--500\,GeV 
for the seven VHE sources and corresponding pulsars.
Green diamonds indicate the positions of \gr\ pulsars and yellow ones the
positions of the other non--\gr\ pulsars.
The positional error circles and extension region of LHAASO, HAWC, and HESS 
	sources are marked with solid and dash circles respectively. 
	All the catalog \fermi-LAT sources are also indicated with their
	catalog names.
\label{fig:tsmapa}}
\end{figure*}

\subsection{PSR J1826$-$1334}
\label{sec:18}
This pulsar is relatively young, with a characteristic age of $\sim 21$\,kyr.
HESS detected largely extended VHE emission with a size of $\sim 1\deg$ 
around this pulsar. Together with 4FGL~J1824.4$-$1350e, the VHE emission 
has been identified as arising from the
PWN of the pulsar \citep{aab+06,kku+18,dgc+19,haa+20}. However, from the
theoretical point of view, the large extension is not easily explained with
the typical PWN modeling (see details in \citealt{kku+18} and 
\citealt{cre+24}).  We included this source as a potential TeV halo case. 

\subsection{PSR J1928+1746}
HAWC detected VHE emission associated with this pulsar \citep{3hwc}, 
named 3HWC J1928$+$178. This radio pulsar \citep{cfl+06} was not found with
an X-ray counterpart in {\it Chandra} and {\it NuSTAR} observations 
\citep{kdp+12,maf+20}. \citet{aaaa+23} re-analyzed the latest HAWC data, 
and two components were revealed in the region of 3HWC~J1928$+$178, 
one with a size of $\sim 0\fdg18$ and another with an extremely large size
of $\sim 1\fdg43$.
We analyzed \fermi-LAT data, but no significant GeV emission was detected 
at the position of the pulsar or in the surrounding region 
(Figure~\ref{fig:tsmapa}).
Considering non-detection of X-ray emission, the age of the pulsar, 
the extendedness of 3HWC~J1928$+$178, and the low energy density compared 
to the local interstellar medium, 
the HAWC source was suggested to be in a transitional phase from a PWN 
to a TeV halo \citep{aaaa+23}. 

LHAASO also detected a source close to PSR~J1928+1746, 
1LHAASO~J1928+1746u, but the extension was only found in WCDA observations, 
not in KM2A ones. In addition, there was a nearby source 1LHAASO J1928+1813u 
that has a large extension of $\sim 0\fdg63$, only detected by KM2A 
(Table~\ref{tab:psr}).
Given the high positional 
coincidence of this source with that of the largely extended component 
identified by HAWC, we considered that these two could be associated with 
each other.
We included 1LHAASO J1928+1813u instead as a candidate TeV halo that is
associated with the pulsar.

\end{document}